\documentclass[preprint,11pt]{elsarticle}
\usepackage{graphicx}
\usepackage{lscape}
\usepackage{upgreek}
\usepackage{caption} 
\usepackage{hyperref}
\usepackage{todonotes}
\usepackage{amssymb}
\usepackage{lineno}
\usepackage{url}
\usepackage{array}
\usepackage{enumitem}
\usepackage{longtable}
\usepackage{multirow}
\usepackage{soul}
\usepackage{float}
\usepackage{color}
\usepackage{pgfplots}
\usepackage{booktabs}

\usepackage{hyperref}
\hypersetup{
pdfproducer={},  
}

\journal{JSS}

\newcommand{\quickwordcount}{%
  \immediate\write18{texcount -1 -sum -merge \jobname.tex > \jobname-words.sum }%
  \input{\jobname-words.sum} words%
}

\begin{document}
\begin{frontmatter}

\title{Technical Debt Prioritization: State of the Art. A Systematic Literature Review}

\author[LUT]{Valentina Lenarduzzi}
\ead{valentina.lenarduzzi@lut.fi}

\author[CHA]{Terese Besker}
\ead{besker@chalmers.se}

\author[TUNI]{Davide Taibi}
\ead{davide.taibi@tuni.fi}

\author[OSLO]{Antonio Martini}
\ead{antonima@ifi.uio.no}

\author[BIC]{Francesca Arcelli Fontana}
\ead{francesca.arcelli@unimib.it}

\address[LUT]{LUT University, Lathi (Finland)}
\address[CHA]{Chalmers University of Technology, G{\"o}teborg (Sweden)}
\address[TUNI]{Tampere University, Tampere (Finland)}
\address[OSLO]{University of Oslo, Oslo (Norway)}
\address[BIC]{University of Milano-Bicocca, Milan (Italy)}

\begin{abstract}
\textit{Background.} Software companies need to manage and refactor Technical Debt issues.
Therefore, it is necessary to understand if and when refactoring of Technical Debt should be prioritized with respect to developing features or fixing bugs. \\
\textit{Objective.} 
The goal of this study is to investigate the existing body of knowledge in software engineering to understand what Technical Debt prioritization approaches have been proposed in research and industry. \\
\textit{Method.} We conducted a Systematic Literature Review of 557 unique papers published until 2019, following a consolidated methodology applied in software engineering. We included 44 primary studies. \\
\textit{Results.} Different approaches have been proposed for Technical Debt prioritization, all having different goals and proposing optimization regarding different criteria. The proposed measures capture only a small part of the plethora of factors used to prioritize Technical Debt qualitatively in practice. We present an impact map of such factors. However, there is a lack of empirical and validated set of tools. 
\\
\textit{Conclusion.} We observed that Technical Debt prioritization research is preliminary and there is no consensus on what the important factors are and how to measure them. Consequently, we cannot consider current research conclusive. In this paper, we therefore outline different directions for necessary future investigations.
\end{abstract}


\end{frontmatter}

\section{Introduction}
\label{Introduction}
Technical Debt (TD) is a metaphor introduced by Ward Cunningham~\cite{Cunningham1992} to represent sub-optimal design or implementation solutions that yield a benefit in the short term but make changes more costly or even impossible in the medium to long term~\cite{DagstuhlReport}.

Software companies need to manage such sub-optimal solutions. The presence of TD is inevitable~\cite{MARTINI2015237} and even desirable under some circumstances~\cite{BESKER8530048} for a number of reasons, which may often be related to unpredictable business or environmental forces internal or external to the organization. 

However, just like any other financial debt, every TD has an interest attached, or else an extra cost or negative impact that is generated by the presence of a sub-optimal solution~\cite{Li2015}. When such interest becomes very costly, it can lead to disruptive events, such as development crises~\cite{MARTINI2015237}. The current best practices employed by software companies include keeping TD at bay by avoiding it if the consequences are known or refactoring or rewriting code and other artifacts in order to get rid of the accumulated sub-optimal solutions and their negative impact. 

However, companies cannot afford to avoid or repay all the TD that is generated continuously and may be unknown~\cite{MARTINI2015237}. The main business goals of companies are to continuously deliver value to their customers and to maintain their products. Thus, the activity of refactoring TD usually competes with developing new features and fixing defects: Such activities are often prioritized over repayment of TD ~\cite{MARTINI2015237}. It is therefore of utmost importance to understand when refactoring TD becomes more important than postponing a feature or a bug fix. In other words, it is important to understand how to \textit{prioritize TD with respect to features and bugs}.

In addition, recent studies show how different projects and even different types of TD might be associated with different refactoring costs (principal) and negative impact (interest)~\cite{Besker_2018}. This means that some TDs can be more dangerous than others~\cite{Seaman2012UsingTD, Martini_2016}, and it is therefore important to understand how to \textit{prioritize TD with respect to other TD}.

However, there is no overall study reporting the current state of the art and practice related to how to prioritize TD. Our goal in this paper is to survey the existing body of knowledge in software engineering to understand which approaches have been proposed in research and industry to \textit{prioritize} TD.  

For this reason, we performed a Systematic Literature Review (SLR) on the prioritization of TD.

We conducted an SLR in order to investigate the existing body of knowledge in software engineering to understand how TD is prioritized in software organizations and which research approaches have been proposed.

The main contribution of this paper is a report on the state of the art concerning approaches, factors, measures, and tools used in practice or proposed in research to prioritize TD. 

The paper is structured as follows: In Section~\ref{Background}, we describe the background of this review. In Section~\ref{TDP}, we outline the research methodology adopted in this study. Section~\ref{Results} and Section~\ref{Discussion} present and discuss the obtained results. Finally, in Section~\ref{Threats}, we identify the threats to validity and in Section~\ref{Conclusion} draw the conclusion.
\section{Background}
\label{Background}
In this Section, we will explain the meaning of TD in order to avoid confusion or misunderstandings, and we will report on previously published systematic reviews.

\subsection{Technical Debt}
\label{Technical Debt}
The concept of TD was introduced for the first time in 1992 by Cunningham as \textit{"The debt incurred through the speeding up of software project development which results in a number of deficiencies ending up in high maintenance overheads"}~\cite{Cunningham1992}. In 2013, McConnell~\cite{McConnell2013} refined the definition of TD as \textit{"A design or construction approach that's expedient in the short term but that creates a technical context in which the same work will cost more to do later than it would cost to do now (including increased cost over time)"}. In 2016, Avgeriou et al.~\cite{Avgeriou2016}  defined it as \textit{"A collection of design or implementation constructs that are expedient in the short term, but set up a technical context that can make future changes more costly or impossible. TD presents an actual or contingent liability whose impact is limited to internal system qualities, primarily maintainability and evolvability"}.

Li et al.~\cite{Li2015} conducted a systematic mapping study for understanding the concept of TD and created an overview of the current state of research on managing TD. Based on the selected studies (96), they proposed a classification of ten types of TD at different levels, as reported in Table~\ref{tab:TDDefinition}. Since this classification derives from a recent secondary study and is, according to our knowledge, the most complete one available in the literature, we considered it in our search strategy process (Section~\ref{Strategy}) to define our search terms. 

\begin{table}[H]
\centering
\scriptsize
\caption{Definition of Technical Debt~\cite{Li2015}} 
\label{tab:TDDefinition} 
\begin{tabular}[h]
{@{}p{2.8cm}|p{9.5cm}@{}}
\hline
\textbf{TD Type} & \textbf{Definition} \\ \hline
Requirements TD & "refers to the distance between the optimal requirements specification and the actual system implementation, under domain assumptions and constraints" \\ \hline
Architectural TD & "is caused by architecture decisions that make compromises in some internal quality aspects, such as maintainability"\\ \hline
Design TD & "refers to technical shortcuts that are taken in detailed design"\\ \hline
Code TD & "is the poorly written code that violates best coding practices or coding rules. Examples include code duplication and over- complex code"\\ \hline
Test TD & "refers to shortcuts taken in testing. An example is lack of tests (e.g., unit tests, integration tests, and acceptance tests)"\\ \hline
Build TD & "refers to flaws in a software system, in its build system, or in its build process that make the build overly complex and difficult"\\ \hline
Documentation TD & "refers to insufficient, incomplete, or outdated documentation in any aspect of software development. Examples include out-of-date architecture documentation and lack of code comments"\\ \hline
Infrastructure TD & "refers to a sub-optimal configuration of development-related processes, technologies, supporting tools, etc. Such a sub-optimal configuration negatively affects the team's ability to produce a quality product"\\ \hline
Versioning TD & "refers to the problems in source code versioning, such as unnecessary code forks"\\ \hline
Defect TD & "refers to defects, bugs, or failures found in software systems"\\ \hline
\end{tabular}
\end{table}

\subsection{Previous SLR$_s$}
\label{PreviousSLR}
In this Section, we briefly report on previous systematic reviews (Systematic Mapping Studies and Systematic Literature Reviews) available in the source engines, showing their main goals in Table \ref{tab:SLR}).
We present the studies in chronological order in order to show the research evolution regarding TD. 
The first systematic review was published in 2012~\cite{Tom2012} and the last ones, to the best of our knowledge, in 2018~\cite{BeskerSLR2018},\cite{Rios2018}.

Tom et al. \cite{Tom2012} exploited an exploratory case study technique that involves a multivocal literature review, supplemented by interviews with software practitioners and academics, in order to establish the boundaries of the TD phenomenon. As a result, they created a theoretical framework that provides a holistic view of TD, comprising a set of TD dimensions, attributes, precedents, and outcomes. The framework provides  a useful approach to understanding the overall phenomenon of TD for practical purposes. 

Li et al.~\cite{Li2015} investigated TD management (TDM), providing a classification of TD concepts and presenting the current state of research on TDM. They considered publications between 1992 and 2013, ultimately selecting 94 studies. 
The results showed a need for empirical studies with high-quality evidence on the TDM process, application of TDM approaches in industrial contexts, and tools for managing the different TD types during the TDM process. 

Ampatzoglou et al.~\cite{AMPATZOGLOUSLR2015} analyzed research efforts regarding TD, focusing on financial aspects underlying software engineering concepts. They considered publications until 2015, selecting 69 studies. 
The results provide a glossary of terms and a classification scheme for financial approaches to be applied for managing TD. 
Moreover, they discovered that a clear mapping between financial and software engineering concepts is lacking.

Ribeiro et al.~\cite{RibeiroSLR2016} evaluated the appropriate time for paying a TD item and how to apply decision-making criteria to balance the short-term benefits against long-term costs. They considered publications until 2016, selecting 38 studies. 
They identified 14 decision-making criteria that can be used by development teams to prioritize the payment of TD items and a list of types of debt related to the criteria.

Alves et al.~\cite{AlvesSLR2016} investigated what strategies have been proposed to identify and manage TD in software projects, considering publications between 2010 and 2014 and selecting 100 studies.
They proposed an initial taxonomy of TD types and provided a list of indicators to identify TD and  management strategies. Moreover, they analyzed the current state on TD, 
highlighting possible research gaps.
The results showed a growing interest of researchers in the TD area. They identified some gaps regarding new indicator proposals and management strategies and tools for controlling TD. Another gap they identified regards empirical studies for validating the proposed strategies.

Fern\'andez-S\'anchez et al.~\cite{FernandezSLR2017} identified the elements needed to manage TD, considering publications until 2017 and selecting 69 studies.
They did not provide a general overview of the TD phenomenon or of the activities for managing TD. The elements were  classified into three groups (basic decision-making factors, cost estimation techniques, practices and techniques for decision-making) and grouped based on  stakeholders’ points of view (engineering, engineering management, and business-organizational management).

Behutiye et al.~\cite{BehutiyeSLR2017} analyzed the state of the art of TD and its causes, consequences, and management strategies in the context of agile software development (ASD).
They considered publications until 2017 and selected 38 studies, finding potential research areas for further investigation. The study highlighted positive interest in TD and ASD and provided some potential categories that can easily lead to TD, such as "focus on quick delivery” and “architectural and design issues”. 

Besker et al.~\cite{BeskerSLR2018} investigated Architectural TD (ATD), synthesizing and compiling research efforts in order to create new knowledge with a specific interest in ATD. They considered publications between 2005 and 2016, selecting 43 studies.
The results showed a lack of guidelines on how to manage ATD successfully in practice and  of an overall process where these activities are fully integrated. 

Rios er al.~\cite{Rios2018} performed a tertiary study based on a set of five research questions and evaluated 13 secondary studies dating from 2012 to March 2018. They evolved a taxonomy of TD types, identified a list of situations in which debt items can be found in software projects, and organized a map representing the state of the art of activities, strategies, and tools for supporting TD management. Their results can help to identify points that still require further investigation in TD research. For example, they found that there are management activities that do not have any type of support tool.

Recently, Khomyakov et al.~\cite{Khomyakov2019} investigated existing tools for the measurement and analysis of TD, focusing on quantitative methods that could also be automated.
They selected 21 papers out of 331 retrieved. Their results show that many new approaches are being defined to measure TD.

\begin{table}[H]
\centering
\scriptsize
\caption{Previous SLR$_s$} 
\label{tab:SLR} 
\begin{tabular}[H]
{@{}p{0.5cm}|p{0.6cm}|p{6cm}@{}}
\hline
\textbf{ID} & \textbf{Year} & \textbf{Goal} \\ \hline
\cite{Tom2012} & 2012 & Understanding the nature of TD \\ \hline
\cite{Li2015} & 2015 & TD management and TD classification   \\ \hline
\cite{AMPATZOGLOUSLR2015} & 2015 & Financial approaches for managing TD \\ \hline
\cite{RibeiroSLR2016}  & 2016 &TD payment prioritization \\ \hline
\cite{AlvesSLR2016} & 2016 &TD management strategies, TD taxonomy \\ \hline
\cite{FernandezSLR2017} & 2017 & TD management elements \\ \hline
\cite{BehutiyeSLR2017} & 2017 & TD in Agile development\\ \hline
\cite{BeskerSLR2018} & 2018 & Managing architectural TD \\ \hline
\cite{Rios2018} & 2018 & TD types, management strategies \\ \hline
\cite{Khomyakov2019} & 2019 & TD tools \\ \hline
\end{tabular}
\end{table}

\section{Methodology}
\label{TDP}
In order to understand the state of the art and the practice on Technical Debt prioritization, we conducted a systematic literature review based on the guidelines defined by Kitchenham et al.~\cite{Kitchenham2007}, \cite{Kitchenham2013}. We also applied the "snowballing" process defined by Wohlin~\cite{Wohlin2014}.

In this Section, we will describe the goal and the research questions (Section~\ref{RQ}) and report our search strategy approach (Section~\ref{Strategy}). Moreover, we performed a quality assessment (Section~\ref{QualityAssessment}) for each included paper and outlined the data extraction and the analysis (Section~\ref{DataExtraction}) of the corresponding data. 

\subsection{Goal and Research Questions}
\label{RQ}
The study goal was to investigate the existing body of knowledge in software engineering to understand how TD is prioritized
in software organizations and what research approaches have been proposed.

Based on our goal, we defined the following research questions (\textbf{RQ$_s$}):

\vspace{2mm}
\begin{tabular}
{@{}p{1cm}p{10.8cm}@{}}
\textbf{RQ$_1$} & Which types of TD have been investigated mostly?\\ \hline 
\textbf{RQ$_2$} & Which prioritization aspects have been proposed? \\
RQ$_{2.1}$ &  Are papers prioritizing TD vs TD or TD vs Features? \\
RQ$_{2.2}$ & Is the prioritization based on a one-shot activity or on a continuous process? \\\hline
\textbf{RQ$_3$} &  Which factors and measures have been considered for TD prioritization? \\\hline
\textbf{RQ$_4$} &  Which tools have been used to prioritize TD? \\
\end{tabular}
\vspace{2mm}

In order to satisfy our goal, we first investigated which types of TD are investigated mostly by researchers and when they should concentrate research efforts in the future (\textbf{RQ$_1$}). Regarding TD types, we adopted the classification proposed by Li et al.~\cite{Li2015} reported in Table~\ref{tab:TDDefinition}. 
Moreover, we characterized how the different TD types are evaluated, highlighting the measures and information. 

The second research question targets how the investigated research papers address the prioritization process of TD, both in terms of different aspects (\textbf{RQ$_2$}), i.e., whether the prioritization process of TD mainly focuses on different TD items or also includes prioritization between TD items and, e.g., the implementation of new features (\textbf{RQ$_{2.1}$}), and of how the prioritization process is described in terms of its periodicity (\textbf{RQ$_{2.2}$}).


Based on the above RQ$_s$, we aimed at identifying a set of factors and measures considered useful during TD prioritization activities (\textbf{RQ$_3$}). Moreover, we aimed at understanding which measures are considered in the prioritization of the main TD components, principal and interest.

We aim to provide a list of existing tools used to evaluate TD in order to depict the current situation in terms of numbers and the maturity of each tool (\textbf{RQ$_4$}). 

\subsection{Search Strategy}
\label{Strategy}
\begin{figure}[H]
\centering
\includegraphics[width=\linewidth]{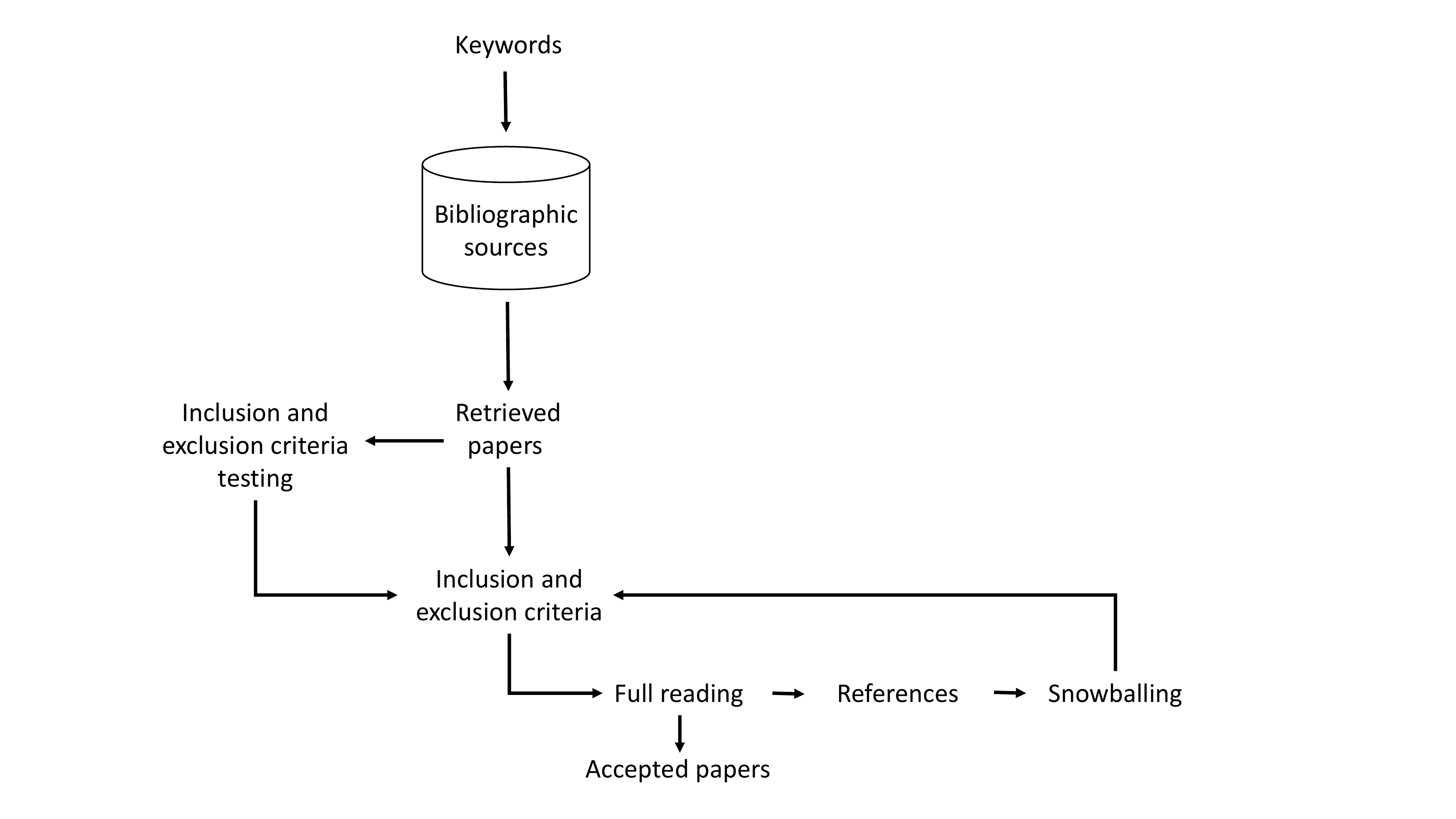}
\caption{The Search and Selection Process}
\label{fig:SelectionProcess}
\end{figure}

The search strategy involves the outline of the most relevant bibliographic sources and search terms, the definition of the inclusion and exclusion criteria, and the selection process relevant for the inclusion decision. Our search strategy is depicted in Figure~\ref{fig:SelectionProcess}.

\textbf{Search terms}. 
In our search string, we included all the terms related to TD proposed by Li et al.~\cite{Li2015} and reported in Table~\ref{tab:TDDefinition} (Section~\ref{Background}).

The search string contained the following search terms: \\
\textbf{("technical debt")OR ("design debt") OR ("architect* debt") OR ("test* debt") OR ("implem* debt") OR ("docum* debt") OR ("requirement debt") OR ("code debt") OR ("Infrastructure debt") OR ("versioning debt") OR ("defect debt") OR ("build debt")}

We used the asterisk character (*) for the second term group in order to capture possible term variations such as plurals and verb conjugations. To increase the likelihood of finding publications addressing TD prioritization, we applied the search string to both title and abstract.

\textbf{Bibliographic sources}. We selected the list of relevant bibliographic sources following the suggestions of Kitchenham and Charters~\cite{Kitchenham2007}, since these sources are recognized as the most representative in the software engineering domain and used in many reviews. The list includes: \textit{ACM Digital Library, IEEEXplore Digital Library, Science Direct, Scopus, Google Scholar, CiteSeer library, Inspec, Springer link}.
Moreover, we performed a manual search on the most important conferences and workshops on Technical Debt, such as the International Conference on Technical Debt (TechDebt).

\textbf{Inclusion and exclusion criteria}. We defined inclusion and exclusion criteria to be applied to the title and abstract (T/A) or to the full text (F) or to both cases (All), as reported in Table~\ref{tab:Criteria}.

\begin{table}[H]
\centering
\scriptsize
\caption{Inclusion and exclusion criteria} 
\label{tab:Criteria} 
\begin{tabular}
{@{}p{1.4cm}|p{9.4cm}|p{0.9cm}@{}}
\hline
\textbf{Criteria} &\textbf{ Assessment Criteria} & \textbf{Step} \\ \hline
\multirow{4}{*}{Inclusion} & Papers that prioritize TD issues & All \\ \cline{2-3}
& Papers that report the criteria of removal\&refactoring\&remediation of TD issues regarding any aspect (financial, maintenance, performance, readability, ...) & All\\ \cline{2-3}
& Papers that compare TD issues & All\\ \cline{2-3}
& Papers that empirically validated/elicited the results & F \\ \hline
\multirow{9}{*}{Exclusion} & Papers not fully written in English  & T/A \\ \cline{2-3}
& Papers not peer-reviewed (i.e., blog, forum ...)\\ \cline{2-3}
& Duplicate papers (only consider the most recent version)& T/A \\ \cline{2-3}
& Position papers and work plans (i.e., papers that do not report results) & T/A \\ \cline{2-3}
& Publications  where  the  full  paper  cannot be  located  (i.e., if database used does not have access to the full text of the publication) & T/A \\ \cline{2-3}
& Publications that only mention prioritization of TD in an introductory statement and do not fully or partly focus on it & All\\ \cline{2-3}
& Only the latest version of the papers (e.g., journal papers that extend conference papers are excluded if they refer to the same dataset)& All\\ \hline
\end{tabular}
\end{table}

\textbf{Search and selection process}. The search was conducted in December 2019 and included all the publications available until this period. The application of the searching terms returned 557 unique papers. 

\textit{Testing the applicability of inclusion and exclusion criteria:} Before applying the inclusion and exclusion criteria, we tested their applicability~\cite{Kitchenham2013} on a subset of ten papers (assigned to all the authors) randomly selected from the papers retrieved.

\textit{Applying inclusion and exclusion criteria to title and abstract:} We applied the refined criteria to the remaining 547 papers. Each paper was read by two authors; in the case of disagreement, a third author was involved in the discussion to clear up any such disagreement.
For 29 papers, we involved a third author. Out of the 557 initial papers, we included 116 based on title and abstract.

\textit{Full reading:} 
We fully read the 116 papers included by title and abstract, applying the criteria defined in Table~\ref{tab:Criteria} and assigning each one to two authors. We involved a third author for six papers to reach a final decision. Based on this step, we selected 49 papers as possibly relevant contributions.

\textit{Snowballing:} We performed the snowballing process~\cite{Wohlin2014}, considering all the references presented in the retrieved papers and evaluating all the papers referencing the retrieved ones, which resulted in one additional relevant paper. We applied the same process as for the retrieved papers. The
snowballing search was conducted in December 2019. We identified only 11 potential papers, but only one of these was included in order to compose the final set of publications.

Based on the search and selection process, we retrieved a total of 50 papers for the review, as reported in Table~\ref{tab:SelectionResults}. 

\subsection{Quality Assessment}
\label{QualityAssessment}
Before proceeding with the review, we checked whether the quality of the selected papers was sufficient to support our goal and whether the quality of each paper reached a certain quality level. We performed this step according to the protocol proposed by Dyb{\aa} and Dings{\o}yr~\cite{Dyba2008}. 
To evaluate the selected papers, we prepared a checklist (Table~\ref{tab:QA}) with a  set of specific questions. We ranked each answer, assigning a score on a five-point Likert scale (0=poor, 4=excellent). A paper satisfied the quality assessment criteria if it achieved a rating higher than (or equal to) 2.

\begin{table}[H]
\centering
\scriptsize
\caption{Quality Assessment Criteria} 
\label{tab:QA} 
\begin{tabular}
{@{}p{0.8cm}|p{8.2cm}|p{2.7cm}@{}}
\hline
\textbf{QA$_s$} & \textbf{Quality Assessment Criteria (QA)}& \textbf{Response Scale} \\ \hline
QA$_1$&Is the paper based on research (or is it merely a "lessons learned" report based on expert opinion)? \\\cline{1-2}
QA$_2$&Is there a clear statement of the aims of the research?\\\cline{1-2}
QA$_3$&Is there an adequate description of the context in which the research was carried out?\\\cline{1-2}
QA$_4$&Was the research design appropriate to address the aims of the research? & Excellent = 4\\\cline{1-2}
QA$_5$&Was the recruitment strategy appropriate for the aims of the research? &  Very Good=3  \\\cline{1-2}
QA$_6$&Was there a control group with which to compare treatments? & Good=2 \\\cline{1-2}
QA$_7$&Was the data collected in a way that addressed the research issue? & Fair=1 \\\cline{1-2}
QA$_8$&Was the data analysis sufficiently rigorous? & Poor=0\\\cline{1-2}
QA$_9$&Has the relationship between researcher and participants been considered to an adequate degree?\\\cline{1-2}
QA$_{10}$&Is there a clear statement of findings?\\\cline{1-2}
QA$_{11}$&Is the study of value for research or practice?\\\hline
\end{tabular}
\end{table}

Among the 50 papers included in the review from the search and selection process, only 44 fulfilled the quality assessment criteria, as reported in Table~\ref{tab:SelectionResults}.

\begin{table}[H]
\centering
\scriptsize
\caption{Results of search and selection and application of quality assessment criteria} 
\label{tab:SelectionResults} 
\begin{tabular}
{@{}p{7.5cm}|p{2cm}@{}}
\hline
\textbf{Step} & \textbf{\# Papers}  \\ \hline
Retrieval from bibliographic sources (unique papers) & 557 \\ \hline
Reading by title and abstract& 439 rejected\\ \hline
Full reading  & 68 rejected\\ \hline
Backward and forward snowballing & 1 \\ \hline
Papers identified & 50 \\ \hline
Quality assessment & 6 rejected \\ \hline
\textbf{primary studies} & \textbf{44} \\ \hline
\end{tabular}
\end{table}

In Table~\ref{tab:SelectedPapers}, we list the 44 papers included in the review (Appendix A reports the details for each paper). The detailed references of all the 44 primary studies is reported in Appendix A.

{\scriptsize
\setlength{\tabcolsep}{6pt}
\begin{longtable}{p{0.8cm}|p{6.7cm}|p{2.7cm}|p{0.7cm}}
\caption{The Selected Papers} 
\label{tab:SelectedPapers} \\ 

\toprule
\textbf{id} & \textbf{Title} & \textbf{Authors} & \textbf{Year} \\ 

\midrule
\endfirsthead
 
\multicolumn{4}{l}{Table \ref{tab:SelectedPapers} continued from previous page} \\ 

\toprule
\textbf{id} & \textbf{Title} & \textbf{Authors} & \textbf{Year}  \\
\midrule
\endhead

\multicolumn{4}{r}{Continued on next page} \\
\endfoot

\bottomrule
\endlastfoot
\ref{SP1}	&	An empirical model of technical debt and interest	&	Nugroho, A. et al.	&	2011	\\ \hline
\ref{SP2}	&	Investigating the impact of design debt on software quality	&	Zazworka, N. et al.	&	2011	\\ \hline
\ref{SP3}	&	Prioritizing design debt investment opportunities	&	Zazworka, N. et al.	&	2011	\\ \hline
\ref{SP4}	&	Estimating the principal of an application's technical debt	&	Curtis, B. et al.	&	2012	\\ \hline
\ref{SP5}	&	Investigating the impact of code smells debt on quality code evaluation	&	Arcelli Fontana, F. et al.	&	2012	\\ \hline
\ref{SP6} & Using technical debt data in decision making: Potential decision approaches & Seaman, C. et al. & 2012\\ \hline
\ref{SP7} & Defining the decision factors for managing defects: A technical debt perspective & Snipes, W. et al. &2012\\ \hline
\ref{SP8} & A formal approach to technical debt decision making & Schmid, K. &2013\\ \hline
\ref{SP9} & Challenges to and Solutions for Refactoring Adoption: An Industrial Perspective	&	Sharma, T. et al.	&	2015	\\ \hline
\ref{SP10} 	&	Investigating Architectural Technical Debt accumulation and refactoring over time: A multiple-case study	&	Martini, A. et al. 	&	2015	\\ \hline
\ref{SP11} &	On the use of time series and search based software engineering for refactoring recommendation	&	Wang, H. et al.	&	2015	\\ \hline
\ref{SP12} 	&	Towards Prioritizing Architecture Technical Debt: Information Needs of Architects and Product Owners	&	Martini, A. and Bosch, J.	&	2015	\\ \hline
\ref{SP13} 	&	Validating and prioritizing quality rules for managing technical debt: An industrial case study	&	Falessi, D. and Voegele, A.	&	2015	\\ \hline
\ref{SP14} 	&	Developing processes to increase technical debt visibility and manageability – An action research study in industry	&	Yli-Huumo, J. et al.	&	2016	\\ \hline
\ref{SP16}	&	How do software development teams manage technical debt? – An empirical study	&	Yli-Huumo, J. et al.	&	2016	\\ \hline
\ref{SP17}	&	Identifying and quantifying architectural debt	&	Xiao, L. et al.	&	2016	\\ \hline
\ref{SP18}	&	JSpIRIT: A flexible tool for the analysis of code smells	&	Vidal, S. et al.	&	2016	\\ \hline
\ref{SP19}	&	Minimizing refactoring effort through prioritization of classes based on historical, architectural and code smell information	&	Choudhary, A. and Singh, P.	&	2016	\\ \hline
\ref{SP20}	&	Pragmatic approach for managing technical debt in legacy software project	&	Gupta, R.K. et al.	&	2016	\\ \hline
\ref{SP21}	&	Technical debt prioritization using predictive analytics	&	Codabux, Z. and Williams, B.J.	&	2016	\\ \hline
\ref{SP22}	& Technical Debt Management with Genetic Algorithms	& Vathsavayi, S. H.  and Systa, K.	& 2016 \\ \hline
\ref{SP23}	&	A Heuristic for Estimating the Impact of Lingering Defects: Can Debt Analogy Be Used as a Metric?	&	Akbarinasaji, S. et al.	&	2017	\\ \hline
\ref{SP24}	&	A strategy based on multiple decision criteria to support technical debt management	&	Ribeiro, L.F. et al.	&	2017	\\ \hline
\ref{SP25}	&	An empirical assessment of technical debt practices in industry	&	Codabux, Z. et al.	&	2017	\\ \hline
\ref{SP26}	&	Assessing code smell interest probability: A case study	&	Charalampidou, S. et al.	&	2017	\\ \hline
\ref{SP27}	&	Impact of architectural technical debt on daily software development work - A survey of software practitioners	&	Besker, T. et al.	&	2017	\\ \hline
\ref{SP28}	&	Investigating the identification of technical debt through code comment analysis	&	de Freitas Farias, M.A. et al.	&	2017	\\ \hline
\ref{SP29}	&	Lessons learned from the ProDebt research project on planning technical debt strategically	&	Ciolkowski, M. et al.	&	2017	\\ \hline
\ref{SP30}	&	Looking for Peace of Mind? Manage Your (Technical) Debt: An Exploratory Field Study	&	Ghanbari, H. et al.	&	2017	\\ \hline
\ref{SP31}	&	Revealing social debt with the CAFFEA framework: An antidote to architectural debt	&	Martini, A., Bosch, J.	&	2017	\\ \hline
\ref{SP32}&	Technical debt interest assessment: From issues to project	&	Martini, A. et al. 	&	2017	\\ \hline
\ref{SP33}	&	The magnificent seven: Towards a systematic estimation of technical debt interest	&	Martini, A., Bosch, J.	&	2017	\\ \hline
\ref{SP34}	&	The pricey bill of Technical Debt - When and by whom will it be paid?	&	Besker, T. et al.	&	2017	\\ \hline
\ref{SP35}&	A semi-automated framework for the identification and estimation of Architectural Technical Debt: A comparative case-study on the modularization of a software component	&	Martini, A. et al. 	&	2018	\\ \hline
\ref{SP36}	&	Early evaluation of technical debt impact on maintainability	&	Conejero, J.M. et al.	&	2018	\\ \hline
\ref{SP37}	&	Technical Debt tracking: Current state of practice: A survey and multiple case study in 15 large organizations	&	Martini, A. et al. 	&	2018	\\ \hline
\ref{SP38}	&	Identifying and Prioritizing Architectural Debt Through Architectural Smells: A Case Study in a Large Software Company	&	Martini, A. et al. 	&	2018	\\ \hline
\ref{SP39}	&	Prioritize technical debt in large-scale systems using codescene	&	Tornhill A. 	&	2018	\\ \hline
\ref{SP40}	&	Prioritizing technical debt in database normalization using portfolio theory and data quality metrics	& Albarak M. and Bahsoon R.		&	2018	\\ \hline
\ref{SP41}	&	Towards a Technical Debt Management Framework based on Cost-Benefit Analysi	& Firdaus H.M. and Lichter H. 	&	2018	\\ \hline
\ref{SP42}	&	Design debt prioritization: a design best practice-based approach	&	Pl{\"o}sch R. et al. 	&	2018	\\ \hline
\ref{SP43}	&	Aligning Technical Debt Prioritization with Business Objectives: A Multiple-Case Study	&	Rebouças R. et al.	&	2018	\\ \hline
\ref{SP44}	&	Technical Debt Prioritization:  A Search-Based Approach &	Alfayez R. and Boehm B.	&	2019	\\ \hline
\end{longtable}
}

\subsection{Data Extraction}
\label{DataExtraction}
We extracted data from the 43 primary studies (PSs) that satisfied the quality assessment criteria. 
The context of each PS is explained in terms of: Context Data, Process Data, and Outcome Data, as reported in Table~\ref{tab:DataExtraction}.

\textbf{Context Data} is necessary to outline the context of each PS in terms of the type of evaluated TD, according to the list proposed by~\cite{Li2015}. We also extracted data regarding the projects considered in the study, such as number of projects, project size, 
and programming languages. Moreover, we collected information about the process phase where the TD is evaluated.

\textbf{Process Data} explain the process adopted to evaluate and prioritize TD issues. We collected data on the type of process (single activity or continuous process, proactive or reactive) and the type of analysis, distinguishing between qualitative, quantitative, and mixed evaluation approaches.
We also retrieved information about the frameworks and tools adopted to evaluate and prioritize TD issues. This data is exclusively based on what is reported in the papers, without any kind of personal interpretation. 

\textbf{Outcome Data} identifies the criteria of removal/refactoring/remediation of TD issues. Moreover, we extracted the measures and factors used to assess the prioritization of a TD issue and which of these are suggested during the prioritization process.

\begin{table}[H]
\centering
\scriptsize
\caption{Data Extraction} 
\label{tab:DataExtraction} 
\begin{tabular}
{@{}p{2cm}|p{10cm}@{}}
\hline
\textbf{Category} & \textbf{Type} \\ \hline
\multirow{3}{*}{Context Data} & Technical Debt type (according to \cite{Li2015}) \\ \cline{2-2}
& Analyzed project (\# of projects, and programming languages)   \\ \cline{2-2}
& Process phase (i.e., maintainability, changeability, etc.)\\ \hline
\multirow{3}{*}{Process Data} & Analysis type (qualitative, quantitative, or mixed evaluation approach)\\ \cline{2-2}
& Frameworks and tools adopted\\ \cline{2-2}
& Process type (single activity or continuous process, proactive or reactive) \\ \hline
\multirow{2}{*}{Outcome Data} & Criteria of removal/refactoring/remediation of TD issues\\ \cline{2-2}
& Measures and factors used to assess the prioritization of a TD issue\\ \hline
\end{tabular}
\end{table}

\subsection{Replicability}
\label{Replicability}
In order to allow replication and extension of our work by other researchers, we prepared a replication package\footnote{http://www.taibi.it/raw-data/JSS\_TD\_2019.zip (The raw data will be moved to a permanent repository (Mendeley Data) in case of acceptance of this paper).} for this study with the complete results obtained.

\section{Results}
\label{Results}

\subsection{Overview of the Primary Studies}
\label{Overview}
\begin{figure}[H]
\centering
\begin{tikzpicture}
\begin{axis}[
label style={font=\footnotesize},
every axis/.append style={font=\footnotesize},
height=2.6in,
width=0.8\linewidth,
xtick={2011,2012,2013,2014,2015,2016,2017,2018},
xticklabels={2011,2012,2013,2014,2015,2016,2017,2018},
ytick={0,1,2,3,4,5,6,7,8,9,10,11,12},
xlabel=Year,
ylabel=Number of Papers,
bar width=15,
ybar
]
\addplot coordinates {(2011,3) (2012,3) (2013,1) (2015,5) (2016,10) (2017,12) (2018,8)};
\end{axis}
\end{tikzpicture}
\caption{Paper Distribution by Year}
\label{fig:PaperYear}
\end{figure}
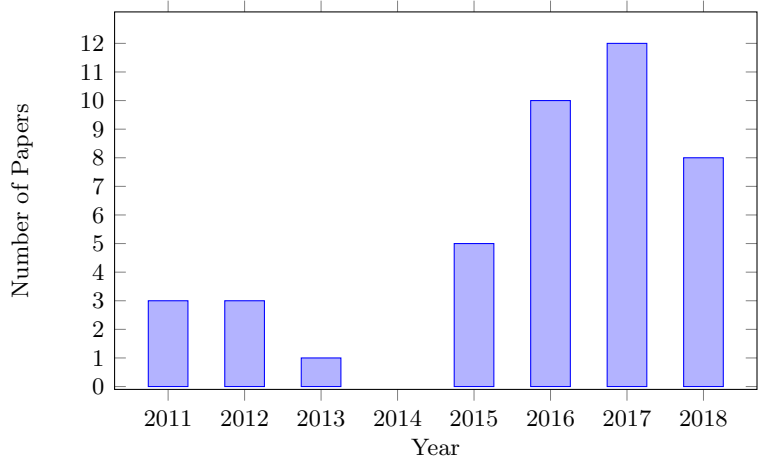

Based on the adopted selection process, we identified 39 primary studies (PSs) as listed in Table~\ref{tab:SelectedPapers}. We illustrate the distribution by year in Figure~\ref{fig:PaperYear}.

The first three relevant papers on TD prioritization were published in 2011. In the next two years, between 2012 and 2014, only three papers were published. 
From 2015, the publication trend increased a lot (5 papers), experiencing a considerable increase in 2016, 2017, and 2018 with 10, 12, and 8 papers, respectively. 

The selected PSs are published in 22 different sources, including 6 journals and 15 conferences and workshops. Specifically, the journal publication sources are: (2 papers) Information and Software Technology (IST), (2 papers) Journal of Systems and Software (JSS), (2 papers) IEEE Software, (1 paper) Empirical Software Engineering Journal (EMSE), (1 paper) Journal of Software: Evolution and Process (JSEP), (1 paper) Science of Computer Programming.

Regarding conferences and workshops, the numbers are: (10 papers) International Conference on Technical Debt (TechDebt) (former Workshop on Managing Technical Debt (MTD)), (4 papers) Euromicro Conference on Software Engineering and Advanced Applications (SEAA),(3 papers) International Conference on Agile Software Development (XP), (2 papers) International Conference on Product-Focused Software Process Improvement (PROFES), (2 papers) International Conference on Software Engineering (ICSE), (1 paper) International Conference on Management of Digital Eco Systems (MEDES), (1 paper) International Conference on Services Computing (SCCC),  (1 paper) International Workshop on Quantitative Approaches to Software Quality (QuASoQ), (1 paper) International Workshop on Emerging Trends in Software Metrics (WETSoM),  (1 paper) International Conference on Enterprise Information Systems (ICEIS), (1 paper) International Symposium on Empirical Software  Engineering and Measurement (ESEM),  (1 paper) International Conference On Software Architecture Workshop (ICSAW), (1 paper) International Conference on Software Maintenance and Evolution (ICSME), (1 paper) International Conference on Quality of Software  architectures (QoSA), (1 paper) International Conference on Software Engineering Advances (ICSEA). 

\subsubsection{Context Data}
\label{ContextData}
28 PSs (75.67\%) conducted case studies in order to investigate TD issues, analyzing different sets of projects. 24 out 28 PSs report the findings for each analyzed project in terms of projects number, project size, and programming language.

Regarding the number of projects analyzed, the majority of the PSs considered fewer than seven each, with most considering only one project.
We identified three papers that took into account as context a huge number of projects, such as \ref{SP4} with 700 projects, \ref{SP1} with 44 projects, and \ref{SP5} with 12 projects. Only 11 PSs report on the programming language of the project(s), with Java, C\#, and C++ being the most common ones.  

The remaining papers investigated TD issues based on surveys among different practitioners. 

TD issues were mainly (48.64\%) investigated with a focus on the maintainability process. The remaining PSs took into account different process phases such as defectively or changeability.

\subsection{RQ$_1$ Which types of TD have been investigated mostly?}
\label{RQ1}
\begin{figure}[H]
\centering
\includegraphics[trim={2cm 0.2cm 2cm 0.05cm},clip, width=0.55\linewidth]{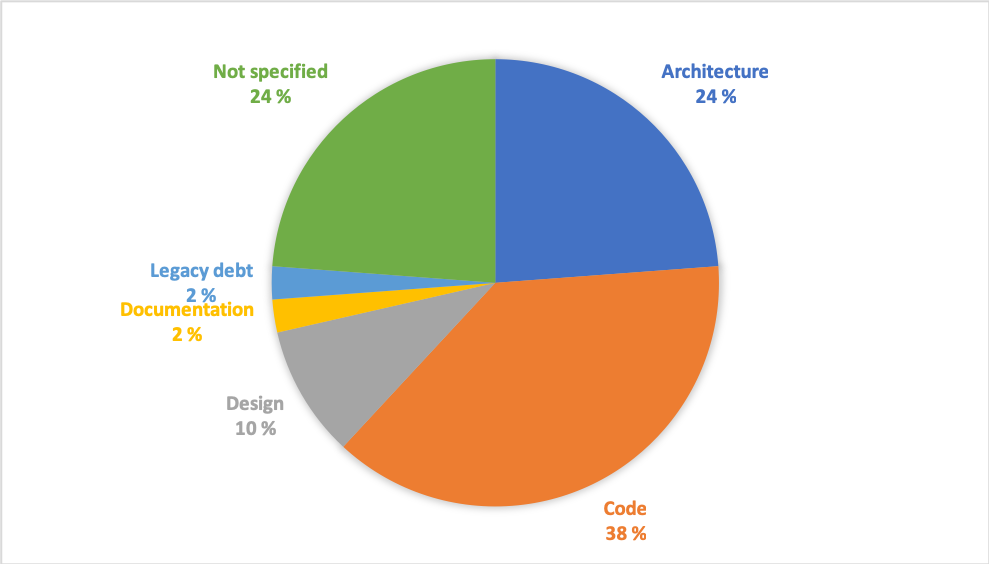}
\caption{Types of TD}
\label{fig:TdTypesPie}
\end{figure}

Considering the TD type reported in Table~\ref{tab:TDDefinition}, the types of TD considered most frequently in the PSs were: Code Debt (38\%), Architectural Debt (24\%), and Design Debt (10\%). Moreover, some PSs (24\%) do not report on issues of any specific TD type, but evaluate TD in general (Figure~\ref{fig:TdTypesPie}). 

Code TD is generally investigated from the point of view of its impact on one - or more than one - software qualities~\ref{SP13}, \ref{SP18}, \ref{SP19}, \ref{SP26}.
Maintainability~\ref{SP4}, \ref{SP5}, \ref{SP11} and maintenance effort~\ref{SP1}, \ref{SP2}, \ref{SP11}, \ref{SP19} are considered most often by the PSs. 
Code debt evaluation is mostly based on code smells~\ref{SP2}, \ref{SP5}, \ref{SP11}, \ref{SP18}, \ref{SP19}, \ref{SP26}.

Other metrics are also considered, such as the time~\ref{SP4}, \ref{SP23} or cost~\ref{SP1} needed to fix a violation, and quality rules~\ref{SP13}.

Some factors related to subjective evaluation such as customer feedback~\ref{SP23} or developers' comments in the code~\ref{SP28} are evaluated less often. 

The approaches mainly involve models that reduce TD by removing or refactoring code smells or other metrics~\ref{SP11},\ref{SP18}. 
These approaches look at the impact on code smells~\ref{SP5}, make a comparison with classes without smells~\ref{SP2}, \ref{SP26}, or rank the code rules~\ref{SP13} perceived as critical by developers. 

Architectural TD is general investigated taking into account the role of architectural smells~\ref{SP17}, \ref{SP19}, \ref{SP20} or complex architectural design~\ref{SP17}, \ref{SP27} which negatively impact software quality~\ref{SP17}, \ref{SP19}, \ref{SP20}. 
Architectural TD is evaluated by measuring the extra maintenance effort for bug fixing~\ref{SP17} or analyzing the bug-proneness~\ref{SP17} of the code.
Another approach combines three different perspectives, such as historical data of the projects, architectural design, and severity of the class prioritizing the refactoring activities~\ref{SP19}.

Architectural design is used to identify high interest in terms of wasted time related to architectural TD~\ref{SP27}, combined with other metrics such as number of files and percentage of complex functions and files~\ref{SP35}. 

Another approach identifies dependencies and social gaps across architecture organization in order to define architectural TD~\ref{SP31}. 

\subsection{RQ$_2$ Which prioritization aspects have been proposed?}
\label{RQ2}
TD prioritization is considered as one of the most important activities when managing TD. The TD prioritization process is used for defining the ordering and/or scheduling of planned refactoring initiatives based on the priority of each identified TD item concerning the impact of the individual items on the software. Several different prioritization aspects have been proposed by researchers in the reviewed publications and a few methods on how to prioritize TD have been developed, but there is no unified approach regarding how the TD prioritization process should be carried out, nor is there a consensus on which aspects to focus on when performing the TD prioritization process. The selection of the prioritization strategy is currently context-dependent in most organizations~\ref{SP21}.

In order to analyze the prioritization aspects presented in the retrieved publications, a thematic analysis approach was used. Thematic analysis is an effective method for identifying, analyzing, and reporting patterns and themes within a searched data scope~\cite{Braun2006}.
The thematic analysis returned mainly five themes illustrating different prioritization aspects. However, one should note that from a software evolution perspective, these aspects can potentially have dependencies and couplings.

Based on the analysis, the different suggested prioritization strategies presented in the reviewed publications are mainly: a) improving software quality, b) increasing software practitioners’ productivity, c) affection on the correctness of the software, d) cost-benefit analysis (CBA) to compare various TD items with respect to low cost and high payoff, or e) a combination of several different approaches.

Studies focusing on internal software quality as a prioritization strategy commonly focus on a quality assessment of the software in order to identify the TD items that cause the highest maintenance costs~\ref{SP1}, ~\ref{SP2}, ~\ref{SP13}, ~\ref{SP19}, ~\ref{SP28}, ~\ref{SP26}, ~\ref{SP4}, ~\ref{SP31}, ~\ref{SP35}, ~\ref{SP41},~\ref{SP44}, together with factors such as remaining product life, debt severity and its impact on future development activities, and current business-related constraints ~\ref{SP3}, ~\ref{SP9}.

Xiao et al.~\ref{SP17} suggest an approach that focuses on architectural TD. It focuses both on locating TD items and on ranking and prioritizing them. Their approach returns the TD items that consume the largest maintenance eﬀort and therefore deserve more attention and higher priority for refactoring

Pl\"{o}sh et al. proposes a TD prioritization approach with a primarily focus on the prioritization of Design debt, and their approach relies on the quantification of design best practices by transferring the identified TD items into a portfolio-matrix~\ref{SP42}. Albarak and Bahsoon further claim that software systems having database tables below fourth normal form are likely to form TD and therefore the  ill-normalized tables should  be prioritized for refactoring~\ref{SP40}.

Other reviewed publications also take the decrease in software practitioners' productivity into consideration when prioritizing TD, since software suffering from architectural TD, for example, slows down development by causing rework ~\ref{SP2}, ~\ref{SP3}.

Also, the effect TD has on the correctness of the software is described as an approach for evaluating different candidate TD items for prioritization~\ref{SP2}. More specifically, Arcelli Fontana, Ferme and Spinelli~\ref{SP5} report that the prioritization of the refactoring of code smells representing design debt can be evaluated by studying the impact of the refactoring of the code smells on different quality metrics,  with the goal is to identify and prioritize ''the most dangerous smell and hence the smell which represents the worst TD''. When prioritizing defect debt, in particular, Akbarinasaji et al.~\ref{SP23} focus their approach on the severity of the debt items (using the categorizations critical, major, normal, and minor) and the duration of bug-ﬁxing time. 

Codabux et al.~\ref{SP21} used a Bayesian approach to build a prediction model for determining the ''TD proneness'' of each TD item using a classification scheme according to the TD proneness probability where the risk of the individual items is assessed. 

Other researchers such as \ref{SP3}, \ref{SP6} use a cost-benefit analysis when prioritizing different TD items, focusing on which refactoring activities should be performed first because they are likely to be inexpensive to implement yet have a significant effect, and which refactoring should be postponed due to high cost and low payoff. The main focus of this approach is on making a lucrative investment in the software, with the output of this analysis being a prioritized list of different TD items ordered by the profitability of the different possible refactoring activities~\ref{SP3}. 

This strategy is echoed by Martini et al.~\ref{SP32}, who state that ''if the interest is (or is going to be) high, the debt is worth being paid. On the contrary, if the interest is not enough to justify the cost of refactoring, there is no reason to ''waste'' resources to refactor the system.''. However, Martini et al.~\ref{SP32} also stress the importance of not only focusing the prioritization decisions on single TD items by assessing each TD item separately, but also understanding the overall impact TD items generally have on the whole project, thus focusing on the overall project goals by evaluating the information holistically. In this approach, Martini et al.~\ref{SP32} also include factors such as the portion of the code affected by the TD, the project size, the roadmap, the positive impact of the TD, the existence of an alternative, and the cultural attitude of the team when prioritizing TD refactoring activities.

Further, Alfayez and Boehm~\ref{SP44} propose an automated search-based approach for prioritizing TD using a multi-objective evolutionary algorithm called MOEA (which is an open-source Java library), having a focus on the repayment of the TD refactoring activity within a speciﬁc cost constraint. 

Borrowing prioritization approaches from other disciplines, such as finance and psychology, Seaman et al.~\ref{SP6} include techniques such as Analytic Hierarchy Process (AHP), the Portfolio method, and the Options approach. The AHP approach involves building a criteria hierarchy, assigning weights and scales to the criteria, and finally performing a series of pairwise comparisons between the alternatives against the various criteria. The goal of using the Portfolio approach is to select those assets that maximize the return on investment or minimize the investment risk. 

Codabux et al.~\ref{SP25} stress the importance of adopting a broader perspective on the prioritization process, focusing on the liability of TD. According to them, decision makers need to think beyond the cost associated with fixing the debt, including estimates of the possible future costs resulting from the decision to ship. The additional costs reflected during the prioritization in terms of liability costs include, e.g., costs for responding to support requests, costs associated with catastrophic failures, etc., and potential litigation costs if service level agreements are violated because of unmanageable debts.

Ribeiro et al.~\ref{SP24} present a multiple decision strategy criteria model using a combination of different prioritization approaches, which can be used during different project phases. Their model focuses on aspects such as, e.g., the severity of the impact the TD items have from a customer perspective on the interest cost of TD, the lifetime of the project's properties, and its possibility of evolution. 

Yet another prioritization process that includes different perspectives is the approach described by Ciolkowski et al.~\ref{SP29}. Their approach focuses on a combination of the overall software quality with a focus on productivity improvement from a future-oriented perspective, using a proactive methodology. 

Gupta et al.~\ref{SP20} use a two-level approach for prioritizing TD. First, the TD items are assessed according to their importance and urgency. In a second step, the TD items' impact on business values and effort is assessed. 

Guo et al.~\ref{SP15} present a TD prioritization approach that ranks customer expectations according to top priority, followed by availability of development resources, the interest of the TD items, the current status of the debt-infected modules, and the impact of the debt on other features. By studying how software practitioners prioritize TD items in practice, Yli-Huumo et al. ~\ref{SP14}, ~\ref{SP16} concluded that their prioritization approach commonly focuses on scalability, business value, use of a feature, and customer effect.

Snipes et al.~\ref{SP7} suggest a TD prioritization approach that includes a combination of factors such as severity, the existence of a workaround, urgency of the refactoring required by customers, refactoring effort, the risk of the proposed refactoring, and the scope of testing required.

Schmid~\ref{SP8} distinguishes between potential and effective TD, where potential TD is any type of sub‐optimal software system, while effective TD refers to issues in the software system that make further development of that system more difficult. This prioritization approach considers aspects such as evolution cost, refactoring cost, and the probability that the predicted evolution path will be realized.

Further, Almeida et al.~\ref{SP43} suggest to also focus on business objectives when prioritization TD in order to support business expectations and goals.  The researcher compared the differences between a technical prioritization and a business-oriented one, and they state that their results show that “taking business priorities into account can change decisions related to technical debt prioritization”. This prioritization aspect is also described to facilitate the argumentation from the technical side and thereby to convince business stakeholders to prioritize what was previously considered pure-technical problems.

Martini and Bosch~\ref{SP33} propose a tool called AnaConDebt to provide assistance during the TD prioritization process. Their tool assesses the severity of the interest for different TD items, with the calculation of the interest being based on an assessment of seven different factors and their growth. The assessed factors are: 1) reduced development speed, 2) bugs related to the TD item, 3) other qualities compromised, 4) other extra costs, 5) frequency of the issue, 6) spread in the system, and 7) users affected. Vidal et al.~\ref{SP18} also propose  a tool called JSpIRIT for specifically prioritizing source-code-related TD, where the TD items are evaluated according to their importance based on different prioritization criteria. The tool calculates a ranking for a set of code smells according to their importance, where the tool can instantiate to prioritize TD items by different criteria. Examples of such criteria are the relevance of the kind of code smells, the history of the system, or different software metrics, among others. Additionally, the developer can use external information to improve the prioritization. 

Yet another reviewed publication~\ref{SP39} suggests performing TD prioritization using a tool called CodeScene, where factors such as how developers work with the code is taken into consideration. The process uses an complexity trend analysis when calculating the indentation-based complexity of the identified TD items and together with a skilled human observer set out the final TD prioritization.

\subsection{RQ$_{2.1}$ Are papers prioritizing TD vs TD or TD vs Features?}
\label{RQ2.1}
Since today’s software companies  face  increasing  pressure  to  deliver  customer  value,  the  balance  between  spending  developer  time, effort,  and  resources  on  implementing  new  features  or spending it on TD remediation activities, on fixing bugs, or on other system improvements become vital. In this study, we limited the scope to studying the balance between prioritizing the implementation of new features or the remediation of existing TD.

To conclude, this research question seeks to address whether the TD prioritization process mainly focuses on the prioritization among different TD items or whether the TD items are described as competing with the implementation of new features or not.

Budget, resources, and available time are important factors in a software project, especially during the prioritization process, since spending time and effort on refactoring activities commonly infers that less time can be spent on implementing new features, for example. This is one of the main reasons why software companies do not always spend additional budget and effort on the refactoring of TD since they commonly have a strong focus on delivering customer-visible features~\ref{SP18}. 

Ciolkowski et al.~\ref{SP29} describe this situation like this: ''The challenge for project managers is to ﬁnd a balance when using the given budget and schedule, either by reducing TD or by adding technical features. This balance is needed to keep time to market for current product releases short and future maintenance costs at an acceptable level.''  Echo this view stating that “Ideally, actionable  refactoring  targets  should  be  prioritized  based  on the  technical  debt  interest  rate  to  balance  the  trade-oﬀs between  improvements,  risk,  and  new features” \ref{SP39}. 

Furthermore, Martini, Bosch and Chaudron~\ref{SP10} state that TD refactoring initiatives usually get low priority compared to the implementation of new features and that TD that is not directly related to the implementation of new features is often postponed. 

Vathsavayi and Systä~\ref{SP22} echo this notion, stating that ''Deciding  whether to spend resources for developing new features or fixing the debt is a challenging task.'' The researchers highlight that software teams need to prioritize new features, bug fixes, and TD refactoring within the same prioritization process.

However, even if the balance between implementing new features and TD refactoring activities is described as important~\ref{SP31}, the papers investigated in this study commonly focus their prioritization approaches on prioritization among different TD items, with the goal being to determine which item should be refactored first. None of the prioritization approaches described in the surveyed publications explicitly addresses how the prioritization between implementing new features and spending time and effort on the refactoring of TD should be carried out. However, the study by Besker et al.~\cite{Besker2019} states that ''the  pressure  of  delivering customer  value  and  meeting  delivery  deadlines  forces  the software  teams  to  down-prioritize  TD  refactorings continuously in favor of implementing new features rapidly''. 

\subsection{RQ$_{2.2}$ Is the prioritization based on a one-shot activity or on a continuous process?}
\label{RQ2.2}
Just as important as prioritizing TD refactoring activities in a project is to describe a management strategy for the prioritization process.

Therefore, this research question focuses on  how the prioritization process is described in the reviewed publications in terms of its periodicity. We distinguish the different approaches in terms of one-shot activities versus part of a continuous process.

Some of the publications reviewed in this study highlight the TD prioritization process in terms of it being a continuous, integrated, and iterative process~\ref{SP16}, ~\ref{SP22}, whereas others stress the importance of prioritizing TD refactoring within each sprint~\ref{SP15}. Choudhary et al.~\ref{SP19} illustrate the prioritization process as being an integral part of the continuous development process by saying ''ideally  software companies try to incorporate refactoring practices as an integral part of their development and maintenance processes''~\ref{SP9}, and ~\ref{SP39} echos this notion stating that “a systematic management of TD and how to reduce it should also be considered important in each release of the development project”. 

Interestingly, however, the rest of the publications reviewed in this study do not give any  explicit  recommendations on  how often  or  in  what  way  the prioritization of TD should be carried out.

\subsection{RQ$_3$ Which factors and measures have been considered for TD prioritization?}
\label{RQ5}

During the prioritization process, six PSs considered both principal and interest (\ref{SP1}, \ref{SP10}, \ref{SP13}, \ref{SP15}, \ref{SP23}, \ref{SP35}), while four PSs considered only interest (\ref{SP13}, \ref{SP17},  \ref{SP27}, \ref{SP34}).  

Principal is calculated as cost~\ref{SP1}, \ref{SP10} or time~\ref{SP1}, \ref{SP4} needed to fix technical quality issues~\ref{SP1} or violations of quality rules~\ref{SP13}. Other factors are also considered, such as page rank or customer feedback~\ref{SP23}. 

Interest is calculated as extra cost spent on maintenance due to technical quality issues~\ref{SP1}, \ref{SP10}, \ref{SP17}, \ref{SP35} or as wasted time related to different activities (management or refactoring)~\ref{SP27}, \ref{SP34}. 

Principal is compared with interest without considering any item for which the benefit does not outweigh the cost~\ref{SP15}. The factors considered are: customer expectations, which have the top priority, followed by availability of development resources, the interest of the TD items, the current status of the debt-infected modules, and the impact of the debt on other features~\ref{SP15}.

In Table~\ref{tab:RQ5}, we present an ''Impact Map'', which highlights the plethora of factors related to the impact (interest) of TD to be considered for prioritization, and their wide variation across studies and projects. In total, we counted 53 unique factors. 

A few of the factors might overlap, although in different papers the factors are calculated differently. For example, ''number of bugs'' and ''ROI (calculated on number of bugs)'' are obviously overlapping factors, although using the sheer number of bugs or the cost of their impact as indicators might give very different results when prioritizing. In other cases, a generic concept of ''interest'' or ''cost'' has been used, although such values were probably implicitly calculated by the researchers or practitioners taking in consideration some of the other 52 remaining factors explicitly mentioned in the other papers. However, given the reported information, there is no way to perform such a mapping. Thus, we report a generic factor, for example ''risk'', as different from all the other specific ones. 

The factors have been grouped into categories, when possible, to help navigate them. First, we mapped the factors to qualities that are mentioned most often in relation with TD. These categories are ''Evolution'', ''Maintenance'', and ''Productivity''. For example, the current working definition of TD explicitly mentions the impact on maintainability and evolvability. Given the emphasis on such qualities, we first grouped the factors according to them. TD impacting other qualities was gathered under ''System Qualities'' (which do not include the former two). Productivity is also usually associated with TD in the form of extra effort spent because of the debt. 

Next, we proceeded to categorize and group the remaining factors according to what aspect of software development the impact is related to. This can be important in order to understand which roles would be hit the most by such impact and what consequences it might have on the prioritization. As an example, TD can have a direct impact on the ''Customer'' factors, so such TD might be considered more important by some organizations in their prioritization. Understanding the impact on ''Business'' factors can also be very useful in a prioritization against features that are prioritized mostly using business concerns. ''Social'' and ''Project'' factors need to be taken into consideration as well, as non-technical aspects of software development.

For some of the factors, it was not possible to find a common category (''Other factors''), or they were only described as high-level factors without additional details (''Not specified''). 

The majority of the papers focus on the impact of TD on maintainability (12). Some papers focus on productivity (7), evolvability (5), and other system qualities (6), while 5 papers consider the customer perspective. 

Only a few papers take into consideration other factors, such business factors (3), social factors (3), project factors (3), and other non-categorized factors (6). In most of these cases (including the customer aspect), the identified factors have been reported in a single paper or two. This highlights either their specificity for a specific context or a lack of focus on these factors in the literature. In both~\ref{SP10} and ~\ref{SP24}, the authors conducted a survey with practitioners to understand which of these factors are most important for developers, architects, and product owners. In most cases, customer and business factors were considered the most important ones. However, only a few papers address such factors when prioritizing TD, so we can conclude that these factors have been overlooked in the literature.

In quite a few studies (8), the interest (impact) of TD has been identified and assessed as generic interest, interest likelihood, risk, severity, or as customizable by the practitioners. Six papers present factors not categorized specifically in the previously mentioned categories and that represent the impact of TD spanning multiple categories or represent a specific aspect not related to these categories.

Eight other papers assume that the impact of TD is associated with the (co-)occurrence of instances of different issues (e.g., code smells) that are considered sub-optimal (''quantity of debt'' in the table). However, the measures used in different papers differ according to the tools used, and the impact of the individual issues is assumed to be the same or was assigned arbitrarily.
Very few papers (4) use an estimate or a measure of the cost of refactoring (principal) in contrast to the impact of TD (interest). This is in contrast with the theoretical approach (\cite{Chatzigeorgiou2015}, \cite{Martini2016AnaconDebt}, \ref{SP8}), according to which TD needs to be prioritized by taking into consideration both the cost of refactoring and the impact.

{\scriptsize
\setlength{\tabcolsep}{6pt}
\begin{longtable}{@{}p{2.3cm}|p{5cm}|p{3.5cm}|p{0.8cm}@{}}
\caption{Impact Map: Factors and measures related to the interest of TD considered when prioritizing (RQ3)} 
\label{tab:RQ5} \\ 

\toprule
\multirow{2}{*}{\textbf{Category}} & \multirow{2}{*}{\textbf{Factors}} & \multicolumn{2}{c}{\textbf{PSs}} \\ \cline{3-4}
& &  \textbf{ID} & \textbf{\textbf{\#}}  \\ \hline

\midrule
\endfirsthead
 
\multicolumn{4}{l}{Table~\ref{tab:RQ5} continued from previous page} \\ \hline

\toprule
\multirow{2}{*}{\textbf{Category}} & \multirow{2}{*}{\textbf{Factors}} & \multicolumn{2}{c}{\textbf{PSs}} \\ \cline{3-4}
& &  \textbf{ID} & \textbf{\textbf{\#}}  \\ \hline
\midrule
\endhead

\multicolumn{4}{r}{Continued on next page} \\
\endfoot

\bottomrule
\endlastfoot
\multirow{6}{*}{\textbf{Business}} 	&	competitive advantage	&	\ref{SP10}	& \multirow{6}{*}{\textbf{3}} 	\\ \cline{2-3}
	&	lead time	&	\ref{SP10}	&	\\ \cline{2-3}
	&	attractiveness for the market	&	\ref{SP10}	& \\ \cline{2-3}
	&	penalties	&	\ref{SP10}	&	\\ \cline{2-3}
	&	feature usage	&	\ref{SP16}	&	\\ \cline{2-3}
	&	business value	&	\ref{SP16}	&	\\ \cline{2-3}
	&	ROI (calculated per bug)	&	\ref{SP20}	&	\\ \hline
\multirow{4}{*}{\textbf{Customer}} 	&	satisfaction	&	\ref{SP12}	& \multirow{4}{*}{\textbf{5}} 	 \\ \cline{2-3}
	&	long-term satisfaction	&	\ref{SP10}	&	 \\ \cline{2-3}
	&	specific customer value	&	\ref{SP10}	& \\ \cline{2-3}
	&	customer expectations	&	\ref{SP13}	&\\ \cline{2-3}
	&	customer effect	&	\ref{SP16}, \ref{SP24}	&		\\ \hline
\multirow{5}{*}{\textbf{Evolution}} 	&	time of impact on evolution (short- or long-term)	&	\ref{SP8}	& \multirow{6}{*}{\textbf{5}} 	\\ \cline{2-3}
	&	risk of critical impact on evolution (possible crisis)	&	\ref{SP8}	&	\\ \cline{2-3}
	&	impact on other features	&	\ref{SP13}, \ref{SP24}	&	\\ \cline{2-3}
	&	impact on upcoming features	&	\ref{SP22}, \ref{SP24}, \ref{SP32}	&	\\ \hline
\multirow{9}{*}{\textbf{Maintenance}} 	&	modifiability	&	\ref{SP2}, \ref{SP18}, \ref{SP26}, \ref{SP28} &	\multirow{9}{*}{\textbf{12}} \\ \cline{2-3}
	&	number of bugs	&	\ref{SP2}, \ref{SP10}, \ref{SP11}, \ref{SP17}, \ref{SP20}, \ref{SP23}, \ref{SP28}, \ref{SP32}, \ref{SP33}, \ref{SP38} &	\\ \cline{2-3}
	&	maintenance cost	&	\ref{SP10}, \ref{SP17}, \ref{SP35} &	\\ \hline
\multirow{6}{*}{\textbf{System Qualities}} &	robustness	&	\ref{SP4}	&	\multirow{6}{*}{\textbf{6}} \\ \cline{2-3}
	&	performance efficiency	&	\ref{SP2}, \ref{SP4}, \ref{SP12}	&	\\ \cline{2-3}
	&	security	&	\ref{SP4}	&	\\ \cline{2-3}
	&	transferability	&	\ref{SP4}	&	\\ \cline{2-3}
	&	scalability	&	\ref{SP16}	&	\\ \cline{2-3}
	&	generic qualities	&	\ref{SP32}, \ref{SP33}, \ref{SP38} &	\\ \hline
\textbf{Quality Debt} 	&	\# of issues or their co-occurrence	&	\ref{SP9}, \ref{SP16}, \ref{SP28}, \ref{SP29}, \ref{SP25}, \ref{SP32}, \ref{SP35}, \ref{SP36}	&	\textbf{8} \\ \hline
	
\multirow{7}{*}{\textbf{Productivity}}  & \% wasted time (effort)		&	\ref{SP27}, \ref{SP32}, \ref{SP33}, \ref{SP34}, \ref{SP35}, \ref{SP38} & \multirow{7}{*}{\textbf{7}} \\ \cline{2-3} 
	&	number of developers working on TD			&	\ref{SP35}\\ \cline{2-3} 
	&	wasted development hours	&	\ref{SP35}\\ \cline{2-3} 
	&	generic effort			&	\ref{SP24}\\ \cline{2-3} 
	&	coding output/effort			&	\ref{SP29}\\ \hline
\multirow{3}{*}{\textbf{Project Factors}} 	&	availability of resources	&	\ref{SP13}	&	\multirow{3}{*}{\textbf{3}}  \\ \cline{2-3}
	&	project size and complexity	&	\ref{SP32}	&	 \\ \cline{2-3}
	&	postponement of bugs	&	\ref{SP23}	&	\\ \hline
\multirow{4}{*}{\textbf{Social Factors}} 	&	developers' morale	&	\ref{SP30}	&\multirow{4}{*}{\textbf{3}} 	\\ \cline{2-3}
	&	social debt	&	\ref{SP31}	&	\\ \cline{2-3}
	&	positive impact of TD	&	\ref{SP32}	&	\\ \cline{2-3}
	&	team culture	&	\ref{SP32}	&	 \\ \hline

\multirow{8}{*}{\textbf{Other Factors}} 	&	contagious debt	&	\ref{SP10}	&	\multirow{9}{*}{\textbf{6}}  \\ \cline{2-3}
	&	existence of TD solution (alternative)	&	\ref{SP32}	&	\\ \cline{2-3}
	&	spread of impact in the system	&	\ref{SP32}, \ref{SP33}, \ref{SP38} 	&\\ \cline{2-3}
	&	number of users affected	&	\ref{SP32}, \ref{SP33}, \ref{SP38} 	&	\\ \cline{2-3}
	&	frequency of negative impact	&	\ref{SP32}, \ref{SP33}, \ref{SP38} 	& \\ \cline{2-3}
	&	kind of smell	&	\ref{SP18}, \ref{SP24} 	&	 \\ \cline{2-3}
	&	history of the system	&	\ref{SP18}	&	 \\ \cline{2-3}
	&	compromise architecture	&	\ref{SP18}	&	 \\ \cline{2-3}
	&	future cost	&	\ref{SP22}	&	\\ \cline{2-3}
	&	user perception	&	\ref{SP24}	&	 \\ \hline
\multirow{6}{*}{\textbf{Not Specified}} 	&	risk	&	\ref{SP10}, \ref{SP25} &\multirow{6}{*}{\textbf{8}} 	 \\ \cline{2-3}
	&	interest likelihood	&	\ref{SP13}, \ref{SP22} &	 \\ \cline{2-3}
	&	interest	&	\ref{SP13}, \ref{SP24}	&	 \\ \cline{2-3}
	&	severity	&	\ref{SP24}, \ref{SP38}	&	 \\ \cline{2-3}
	&	customizable	&	\ref{SP18}, \ref{SP24}, \ref{SP25}, \ref{SP32}, \ref{SP33}, \ref{SP38}	&	 \\ \hline
\end{longtable}
}

\subsection{RQ$_4$ Which tools have been used to prioritize TD?}
\label{RQ6}

As reported in Table~\ref{tab:RQ4}, only 14 papers mentioned the usage of tools for evaluating and prioritizing TD, but only ten of them report information on which tools were used. 
The other studies used a custom-made tool developed for their specific purposes.

\begin{table}[H]
\centering
\footnotesize
\caption{Tool Used when Prioritizing TD (RQ4)} 
\label{tab:RQ4} 
\begin{tabular}
{@{}p{3cm}|p{6.6cm}|p{2cm}@{}}
\hline
\textbf{Tool Name} & \textbf{Tool Link} & \textbf{Paper ID} \\ \hline
 AnaConDebt & \url{https://anacondebt.com} & \ref{SP32}, \ref{SP33}\\ \hline ARCAN ~\cite{Fontana2016ARCAN}, \cite{Fontana2017ARCAN} & \url{http://essere.disco.unimib.it/wiki/arcan} & \ref{SP38} \\ \hline 
CAFFEA & not available & \ref{SP31} \\ \hline 
CAST & \url{https://www.castsoftware.com} & \ref{SP4} \\ \hline 
Coverity & \url{http://www.coverity.com} & \ref{SP20} \\ \hline 
Findbugs & \url{http://findbugs.sourceforge.net} & \ref{SP20}\\ \hline 
Visual Studio FxCopAnalyzer & \url{https://www.nuget.org/packages/Microsoft.CodeAnalysis.FxCopAnalyzers} & \ref{SP20} \\ \hline 
iPlasma & \url{http://loose.cs.upt.ro/index.php?n=Main.IPlasma} & \ref{SP5} \\ \hline 
Jsprit &  \url{https://sites.google.com/site/santiagoavidal} & \ref{SP18} \\ \hline 
Scitool Understand & \url{https://scitools.com} & \ref{SP21}\\ \hline 
SonarQube & \url{https://www.sonarqube.org} & \ref{SP30}\\ \hline 
Codescene & \url{https://codescene.io} &  \ref{SP39}\\ \hline 
\end{tabular}
\end{table}

Out of the aforementioned tools, we can identify ten static analysis tools: \textit{ARCAN}, \textit{CAST}, \textit{Coverity}, \textit{Findbugs}, \textit{Visual Studio FxCopAnalyzer}, \textit{iPlasma}, \textit{Jspirit}, \textit{Scitool Understand}, and \textit{SonarQube}.
\textit{Scitool Understand} analyzes the code and visualizes its architecture.  The remaining ones  detect TD issues such as code or architectural smells, security violations, or others. 
\textit{CAST}, \textit{Coverity}, \textit{Findbugs}, \textit{Studio FxCopAnalyzer}, \textit{Codescene}, and \textit{SonarQube}  are commercial tools commonly used to analyze code compliance against a set of rules.
When the rules are violated, they raise a TD issue. These tools provide the severity of the issues and classify them into different types (e.g., issues that could lead to bugs,  to increased software maintenance effort, or to  security vulnerabilities). Moreover, \textit{CAST} and \textit{SonarQube} also associate a remediation effort (principal), the time needed to remove the TD issue. 
\textit{ARCAN}, \textit{iPlasma}, and \textit{Jspirit} are open-source tools, developed by research teams and aimed at detecting architectural smells (\textit{ARCAN}) and code smells (\textit{iPlasma} and \textit{Jspirit}). 

\textit{AnaConDebt}~\cite{Martini2018} is a management tool based on a TD-enhanced backlog. The backlog allows the creation of TD items and performs TD-specific operations on the created items. In~\ref{SP32} and ~\ref{SP33}, \textit{AnaConDebt} has been used to report and visualize the information on TD manually collected by product managers and developers. 

The \textit{CAFFEA} framework~\cite{Martini2016CAFFEA} identifies organizational roles, where architectural responsibilities are allocated. Moreover, the tool defines the team members and share among them. The framework has been used in~\ref{SP31} to analyze mismatches between the architecture community and the system architecture.

\textit{ARCAN} was used in~\ref{SP38} to detect architectural smells. The TD principal was then investigated by means of a survey in a large company. 

In~\ref{SP30}, developers were asked to discuss the TD issues raised by \textit{SonarQube}. However, there is no information on whether the developers considered the severity or the type of TD issues. 
In~\ref{SP4}, the authors used \textit{CAST} as is to estimate the principal calculated as time to remove all TD issues.

\textit{iPlasma} and \textit{Jspirit} were used in~\ref{SP5} and ~\ref{SP18}, respectively, to detect the number of code smells to be refactored in the systems under investigation. 

\textit{Scitool Understand} was used in~\ref{SP21} to identify architectural issues in the system under investigation. 

The TD issues detected by \textit{Coverity}, \textit{Findbugs}, and \textit{Visual Studio FxCopAnalyser} were used in ~\ref{SP20} for an industrial survey. 



\section{Discussion}
\label{Discussion}
In this Section, we will discuss the  results obtained, outlining some implications for researchers and practitioners working in the TD domain.

Although the TD domain is relatively young compared to other domains such as software testing or software quality, significant contributions have been published in the last ten years and researchers are becoming more and more active (Figure~\ref{fig:PaperYear}).

Among the ten TD types proposed in 2015 by Li et al.~\cite{Li2015} (Table~\ref{tab:TDDefinition}), only Code Debt and Architectural Debt have been considered frequently by researchers (\textbf{RQ$_1$}) in the context of TD prioritization. 

In the study proposed by Li et al.~\cite{Li2015}, Code Debt was the most commonly investigated type of TD, followed by Test Debt. However, other types of TD have also received significant attention. Differently than in~\cite{Li2015}, in our work it emerged that Code Debt and Architectural Debt are by far the most frequently investigated types of debt when considering TD prioritization. 
This could be due to the fact that they are easy to measure, mainly  based on extensions of previous research from other domains, or it may also be due to the fact that they (particularly ATD) are considered as the most harmful and expensive types to manage in software. For example, architectural and code patterns  have been investigated for more than twenty years, even though they were not considered as ''debt''. 

The two most commonly considered types of TD (Code Debt and Architectural Debt) are mainly evaluated by means of architectural or code-level anti-patterns (architectural smells, code smells, or code violations). Moreover, their harmfulness is mainly related to the influence they have on some external quality (e.g., the impact of a specific code smell on maintenance effort). However, their influence is still not clear, since the vast majority of studies do not agree on their harmfulness. Other types of TD should be investigated in the future. We believe that Code Debt is the type investigated most often since it is easy to access the data by mining software repository studies, while other types of debt require other types of studies, including case studies involving developers. We recommend that practitioners should consider the measures identified in this RQ, but should complement them with expert judgement to understand which architectural smells, code smells, or code violations to consider. 

In a software affected by TD, the only significantly effective way to  reduce this TD is to refactor it. This fact stresses the importance of continuously and iteratively prioritizing the identified refactoring tasks and thereby highlights the importance of using an appropriate TD prioritization process. 
Through this study, we have identified several different approaches for prioritizing TD (\textbf{RQ$_2$}, \textbf{RQ$_{2.1}$}, and \textbf{RQ$_{2.2}$}). However, there is no unified approach for this activity, nor is there a consensus on which aspects to focus on when performing the TD prioritization process. 

It is evidently clear from the findings that the prioritization process of TD refactoring can be carried out using different approaches, all having different goals and proposing optimization with regard to different criteria. 

This study has identified five different main approaches that aim to: a) improve software qualities, especially maintainability and evolvability, b) increase software practitioners’ productivity, c) reduce the fault-proneness of the software, d) compare various TD items using cost-benefit analysis (CBA) to understand the convenience of refactoring, and e) combine several different approaches.

This result is of value to both academics and practitioners and illustrates that is it important to first identify the goals of TD prioritization, and thereafter to implement a corresponding TD prioritization approach targeting the identified and specified goals. 

One interesting finding is that the investigated papers usually only compared different TD items during this prioritizing process and more rarely compared the need for implementing a new feature with the refactoring of TD. 

Regarding the characteristics and measures considered during the prioritization process (\textbf{RQ$_3$}), the results so far imply that prioritizing TD is an activity that requires a holistic view of several factors. The systematic assessment of TD requires a wide amount of information, which might change from case to case, and in most cases TD is prioritized without following a standardized approach. Also, the known measures used in a few papers capture only a small part of the factors that are used to prioritize TD (proxy for maintenance costs or productivity). Using only such measures to prioritize TD without considering the full picture of the relevant factors (risks and costs) might consequently result in partial and thus biased prioritization, which in turn could lead to poor business decisions. On the other hand, some of the factors have been reported in a single study conducted in a specific context and might not be relevant in other prioritization cases. 

More studies are necessary in order to obtain better evidence on factors that have been overlooked (for example factors related to customers, business, social, and project aspects). In addition, we need to better understand which factors should be considered in different contexts, and which additional measures should be considered when prioritizing TD. Finally, although a few holistic approaches have been reported (\cite{Martini2016AnaconDebt}, \ref{SP24}, \ref{SP33}), there is a need for a better defined framework and a standardized approach for assessing TD. 

Considering the two main components of TD, only a limited number of papers propose how to evaluate principal and interest. Interest is mainly calculated as extra cost, or as time wasted to  fix TD issues. The reason could be that TD interest is not easy to calculate without access to empirical data from companies. Researchers should design and perform studies to understand the actual interest of existing TD issues. 

The tool support for prioritization activities is very fragmented (\textbf{RQ$_4$}), which highlights the lack of a solid, widely used, and validated set of tools specifically for TD prioritization. Current tools mainly identify TD issues and, in some cases, propose an estimate of the time needed to fix them. However, to the best of our knowledge, no tools calculate the interest due to the postponement of activities. 

Our results can be useful for both researchers and practitioners. 
Researchers should focus on the other types of TD, also considering types of TD that have not been investigated a lot in the last few years. They can also  evaluate approaches, factors, and measures and how to prioritize them. 
Moreover, since the available tools are not fully mature, research activities can focus on empirical validation of existing tools, confirming the usefulness of each measure proposed by each tool. 

Practitioners can benefit from our results by using our impact map to explore/anticipate what kind of impact might occur because of TD. Moreover, they should be careful in selecting tools, not applying only one but considering more than one. 

\subsection{The TD Prioritization Framework}

\begin{figure}[H]
\centering
\includegraphics[width=\linewidth]{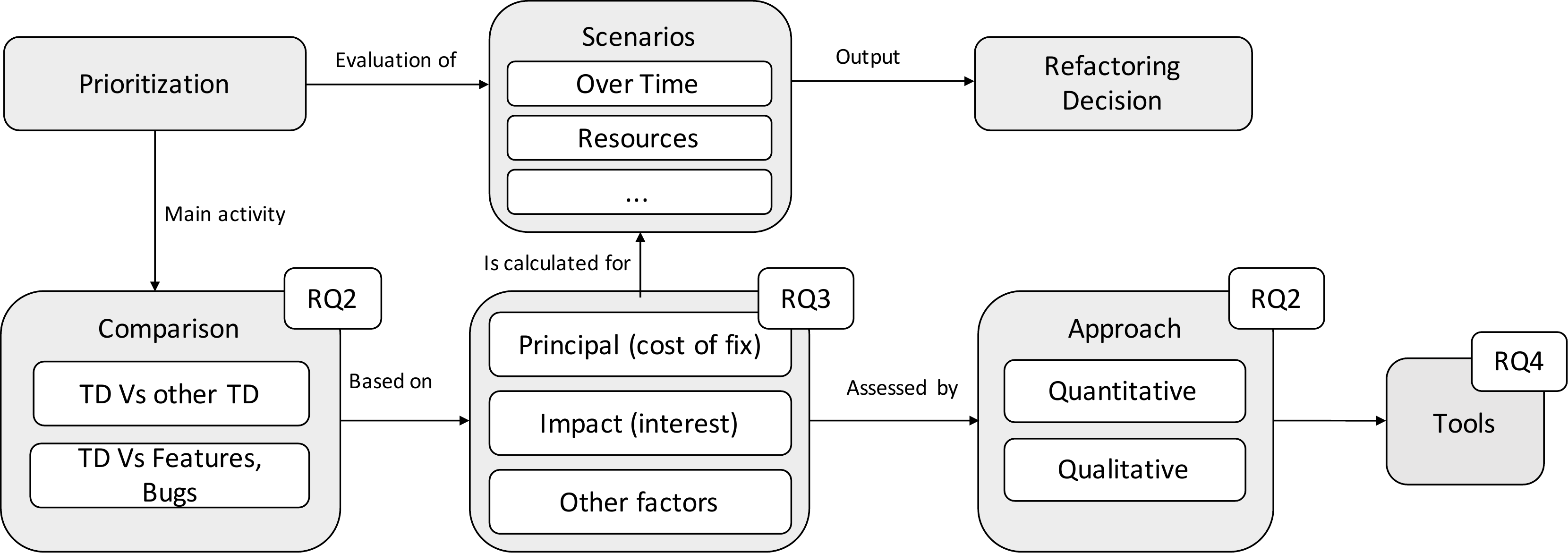}
\caption{The TD Prioritization Framework}
\label{fig:framework}
\end{figure}

Based on our results, we  propose a preliminary framework to  help practitioners during TD prioritization activities as illustrated in Figure~\ref{fig:framework}. 
This framework offers an exploration of different factors that need to be taken into consideration during the TD prioritization process and how these factors relate to each other.

The first step for practitioners is to decide whether they require a prioritization of the refactorings among TD issues or whether they need to prioritize a TD refactoring versus the implementation of new features and bug fixes (results from \textbf{RQ$_2$}). This is because the approaches differ in terms of assessing the impact of TD and assessing the value or the impact of features and bugs. In the former case, the comparison can use the same factors, while in the latter case, it is more probable that the principal and interest of TD need to be compared with feature-oriented factors, for example competitive advantage or cost of delay. 

Once the scope of the comparison is defined, the evaluation of TD should be performed taking into account: 1) the difference in the TD principal (the cost of fixing the issues), 2) the impact (the TD interest), and 3) other factors, including economic and marketing factors (results from \textbf{RQ$_3$}).

The evaluation can be both quantitative and qualitative (\textbf{RQ$_2$}), and in some cases could be supported by tools (\textbf{RQ$_4$}). 
As an example, companies might quantitatively evaluate the presence of Code Debt using tools, but they might also need to perform a qualitative evaluation (e.g., with code reviews) of factors that cannot be measured with tools, for example considering code readability, analyzability, or other quality characteristics. In addition, some tools provide means to calculate the principal of the TD, but practitioners might need to calculate the interest by qualitatively assessing the impact factors.

Moreover, the evaluation should be performed considering different scenarios, including the available resources and the possible evolution of the system. In fact, TD can be quite context-dependent (as we discussed for the impact factors in \textbf{RQ$_3$}), which means that practitioners need to assess it with estimations of future scenarios. For example, in the tool AnaConDebt, practitioners can specify events happening in short-, medium- or long-term scenarios.
The evaluation of the different scenarios should help in making refactoring decisions, for example regarding which refactorings should be performed and which should be postponed. 

As an example of the decision process, a company might consider not implementing a new feature that involves a code section or module that is suffering from TD. This can happen if such TD is estimated to generate high interest in the short-term scenario: In such a case, the interest generated by the TD could overcome the cost of delaying the feature. The practitioners might then decide to refactor the code before implementing the feature.

Let's take a concrete example of how a refactoring decision is made following the steps in the prioritization framework. An architect needs to decide whether to refactor a ''sub-optimal'' interface before more applications accessing it are developed. The main activity of the architect is to evaluate whether to prioritize the refactoring of TD vs. developing new features. Then the architect needs to take into consideration and calculate different factors (principal, interest, and other factors). Without the refactoring, the TD would spread to all the new code (\textit{Contagious debt}, Table \ref{tab:RQ5}). In addition, all the new applications would suffer from the negative impact (interest) generated by interacting with the sub-optimal API (\textit{Spread of impact in the system}, Table \ref{tab:RQ5}). Although delaying the development of the new applications (feature-oriented factor) would imply costs in the short-term scenario, the lead time (Table \ref{tab:RQ5}) for developing new features in the long-term scenario could be reduced as the developers would not pay the interest generated by the sub-optimal interface. If such long-term gain overcomes the cost of delaying the application development, the practitioners should choose to perform the refactoring of the API. In this case, the refactoring decision would be made by evaluating whether, in a future scenario, the cost of avoiding the interest is worth paying the principal. 

The TD prioritization framework can assist practitioners, in combination with the other results presented in this paper (impact map, description of prioritization approaches, and available tools), in reaching a refactoring decision.





\section{Threats to Validity}
\label{Threats}
The results of an SLR may be subject to validity threats, mainly concerning the correctness and completeness of the survey.
In this Section, we will outline some implications for researchers and practitioners working in the TD domain. We have structured this Section as proposed by Wohlin et al.~\cite{Wohlin2012}, including construct, internal, external, and conclusion validity threats.

\subsection{Construct validity}
Construct validity is related to generalization of the result to the concept or theory behind the study execution~\cite{Wohlin2012}. In our case, it is related to the potentially subjective analysis of the selected studies.
As recommended by Kitchenham’s guidelines~\cite{Kitchenham2007}, data extraction was performed independently by two or more researchers and, in case of discrepancies, a third author was involved in the discussion to clear up any disagreement. Moreover, the quality of each selected paper was checked according to the protocol proposed by Dyb{\aa} and Dings{\o}yr~\cite{Dyba2008}. 

\subsection{Internal validity}
Internal validity threats are related to possible wrong conclusions about causal relationships between treatment and outcome~\cite{Wohlin2012}. In the case of secondary studies, internal validity represents how well the findings represent the findings reported in the literature. In order to address these threats, we carefully followed the tactics proposed by~\cite{Kitchenham2007}.

\subsection{External validity}
External validity threats are related to the ability to generalize the result~\cite{Wohlin2012}. In secondary studies, external validity depends on the validity of the selected studies. If the selected studies are not externally valid, the synthesis of its content will not be valid either. In our work, we were not able to evaluate the external validity of all the included studies.

\subsection{Conclusion validity}
Conclusion validity is related to the reliability of the conclusions drawn from the results~\cite{Wohlin2012}. In our case, threats are related to the potential non-inclusion of some studies. In order to mitigate this threat, we carefully applied the search strategy, performing the search in eight digital libraries in conjunction with the snowballing process~\cite{Wohlin2014}, considering all the references presented in the retrieved papers, and evaluating all the papers that reference the retrieved ones, which resulted in one additional relevant paper.  We applied a broad search string, which led to a large set of articles, but enabled us to include more possible results. We defined inclusion and exclusion criteria and applied them first to title and abstract. However, we did not rely exclusively on titles and abstracts to establish whether the work reported evidence on Technical Debt prioritization. Before accepting a paper based on title and abstract, we browsed the full text, again applying our inclusion and exclusion criteria.

\section{Conclusion}
\label{Conclusion}
Software companies need to manage and refactor TD issues since sometimes their presence is  inevitable, due to a number of causes that may be related to unpredictable business or environmental forces internal or external to the organization. 
Moreover, some types of TD can be more dangerous than others. 

Therefore, it is necessary to understand when refactoring TD should be prioritized with respect to implementing features or fixing bugs, or with respect to other types of TD. 

We conducted an SLR in order to investigate the existing body of knowledge in software engineering and gain an understanding of how TD is prioritized
in software organizations and what research approaches have been proposed.  

The SLR process was carried out by following two rigorous approaches.
We included scientific articles indexed by the most important bibliographic sources and selected by a rigorous process. We considered articles published before December 2019.
Our work is based on 38 selected studies, which include data on the state of the art concerning approaches, factors, measures, and tools used in practice or proposed in research to prioritize TD. 

The results of our review show that Code Debt and Architectural Debt are by far the most frequently investigated type of debt when considering TD prioritization, while there is scant evidence about other types of TD such as Test Debt and Requirement Debt. 
The prioritization process of TD refactoring can be carried out using different approaches, all having different goals and proposing optimization with regard to different criteria. However, the identified measures used in a few papers capture only a small part of the factors that are used to prioritize TD.

There is a lack of empirical evidence on measuring principal and interest. Moreover, our results highlight the lack of a solid, validated,  and widely used set of tools specifically for TD prioritization.

In practice, we found that there is a plethora of aspects that need to be considered when prioritizing TD. We presented an impact map of such factors, which can be used as a comprehensive reference regarding which interest might be paid by an organization and how it should be considered. This map can also be used to follow up with further research. 

Future work should focus on the investigation of types of TD that have been investigated less often. Moreover, we are planning to investigate how to systematically evaluate and measure the principal and interest of different types of TD. We also aim at developing a framework to support decision-making related to the prioritization of TD.

\section*{References}

\bibliographystyle{model1-num-names}
\bibliography{TD_Prioritization}
\section*{Appendix A: The Selected Papers} 
\label{The Selected Papers}
{\small
  \begin{enumerate}[labelindent=-5pt,label={[SP}{\arabic*]}]
\item \label{SP1}
A. Nugroho, J. Visser, and T. Kuipers.  An empirical model of technical debt and interest. 2nd Workshop on Managing Technical Debt (MTD '11). pp. 1-8, 2011. 
\item \label{SP2} 
N. Zazworka, M. A. Shaw, F. Shull, and C. Seaman.  Investigating the impact of design debt on software quality. 2nd Workshop on Managing Technical Debt (MTD '11). pp. 17-23, 2011.
\item \label{SP3} 
N. Zazworka, C. Seaman, and F. Shull. Prioritizing design debt investment opportunities. 2nd Workshop on Managing Technical Debt (MTD '11). pp. 39-42, 2011.
\item \label{SP4} 
B. Curtis, J. Sappidi and A. Szynkarski. Estimating the Principal of an Application's Technical Debt. IEEE Software, vol. 29, no. 6, pp. 34-42, 2012.
\item \label{SP5} 
F. Arcelli Fontana, V. Ferme, and S. Spinelli. Investigating the impact of code smells debt on quality code evaluation. Third International Workshop on Managing Technical Debt (MTD '12). pp. 15-22, 2012.
\item \label{SP6} 
C. Seaman et al. Using technical debt data in decision making: Potential decision approaches. Third International Workshop on Managing Technical Debt (MTD'12), pp. 45-48, 2012.
\item \label{SP7} 
W. Snipes, B. Robinson, Y. Guo, and C. Seaman. Defining the decision factors for managing defects: a technical debt perspective.  Third International Workshop on Managing Technical Debt (MTD '12), pp. 54-60, 2012.
\item \label{SP8} 
K. Schmid. A formal approach to technical debt decision making. 9th international ACM Sigsoft conference on Quality of software architectures (QoSA '13), pp/153-162, 2013. 
\item \label{SP9} 
T. Sharma, G. Suryanarayana and G. Samarthyam. Challenges to and Solutions for Refactoring Adoption: An Industrial Perspective. IEEE Software, vol. 32, no. 6, pp. 44-51, 2015.
\item \label{SP10} 
A. Martini, J. Bosch, M. Chaudron. Investigating Architectural Technical Debt accumulation and refactoring over time: A multiple-case study. Information and Software Technology, Volume 67, pp. 237-253, 2015.
\item \label{SP11} 
H. Wang, M. Kessentini, W. Grosky, and H. Meddeb. On the use of time series and search based software engineering for refactoring recommendation. 7th International Conference on Management of computational and collective intElligence in Digital EcoSystems (MEDES '15). pp. 35-42, 2015. 
\item \label{SP12} 
A. Martini and J. Bosch. Towards Prioritizing Architecture Technical Debt: Information Needs of Architects and Product Owners. 41st Euromicro Conference on Software Engineering and Advanced Applications.  pp. 422-429, 2015.
\item \label{SP13} 
D. Falessi and A. Voegele. Validating and prioritizing quality rules for managing technical debt: An industrial case study. 7th International Workshop on Managing Technical Debt (MTD). pp. 41-48, 2015.
\item \label{SP14} 
J. Yli-Huumo, A. Maglyas, K. Smolander, J. Haller and H. T{\"o}rnroos. Developing Processes to Increase Technical Debt Visibility and Manageability - An Action Research Study in Industry. Product-Focused Software Process Improvement. pp. 368-378, 2016. 
\item \label{SP15} 
Y. Guo, R. Oliveira Spínola, and C. Seaman. 2016. Exploring the costs of technical debt management - a case study. Empirical Softw. Engg. Volume 21(1), pp. 159-182, 2016.
\item \label{SP16} 
J. Yli-Huumo, A. Maglyas, and K. Smolander.  How do software development teams manage technical debt? - An empirical study. Journal of  System and  Software, Vol. 120, C, pp. 195-218, 2016.
\item \label{SP17} 
U. Xiao, Yuanfang Cai, Rick Kazman, Ran Mo, and Qiong Feng. Identifying and quantifying architectural debt. 38th International Conference on Software Engineering (ICSE '16), pp. 488-498, 2016.
\item \label{SP18} 
S. Vidal, H. Vazquez, J. A. Diaz-Pace, C. Marcos, A. Garcia and W. Oizumi. JSpIRIT: a flexible tool for the analysis of code smells. 34th International Conference of the Chilean Computer Science Society (SCCC), pp. 1-6, 2015.
\item \label{SP19} 
A. Choudhary and P. Singh. Minimizing Refactoring Effort through Prioritization of Classes based on Historical, Architectural and Code Smell Information. QuASoQ/TDA@APSEC, 2016.
\item \label{SP20} 
R.K. Gupta, P. Manikreddy, S. Naik, and K. Arya.  Pragmatic Approach for Managing Technical Debt in Legacy Software Project. 9th India Software Engineering Conference (ISEC '16), pp. 170-176, 2016.
\item \label{SP21} 
Z. Codabux and B. J. Williams. Technical debt prioritization using predictive analytics. 38th International Conference on Software Engineering Companion (ICSE '16), pp. 704-706, 2016.
\item \label{SP22}
S. H. Vathsavayi and K. Systä. Technical Debt Management with Genetic Algorithms. 42th Euromicro Conference on Software Engineering and Advanced Applications (SEAA), Limassol, pp. 50-53, 2016.
\item \label{SP23} 
S. Akbarinasaji, A. Bener and A. Neal. A Heuristic for Estimating the Impact of Lingering Defects: Can Debt Analogy Be Used as a Metric?. 8th Workshop on Emerging Trends in Software Metrics (WETSoM), pp. 36-42, 2017.
\item \label{SP24}
L. Ferrera Ribeiro, N. S. R. Alves, M. G. d. M. Neto and R. O. Spínola. A Strategy Based on Multiple Decision Criteria to Support Technical Debt Management. 43rd Euromicro Conference on Software Engineering and Advanced Applications (SEAA), pp. 334-341, 2017.
\item \label{SP25}
Z. Codabux, B. Williams, G. Bradshaw and M. Cantor. An empirical assessment of technical debt practices in industry. Journal of Software: Evolution and Process. Vol. 29, 2017.
\item \label{SP26} 
S. Charalampidou, A.  Ampatzoglou, A. Chatzigeorgiou, and P. Avgeriou. Assessing code smell interest probability: a case study. XP2017 Scientific Workshops (XP '17), Article 5, 8 pages, 2017.
\item \label{SP27} 
T. Besker, A. Martini and J. Bosch. Impact of Architectural Technical Debt on Daily Software Development Work — A Survey of Software Practitioners. 43rd Euromicro Conference on Software Engineering and Advanced Applications (SEAA), pp. 278-287, 2017.
\item \label{SP28} 
M. Farias, J. Amâncio Santos, M. Kalinowski, M. Mendonça and R. Spínola, Rodrigo. Investigating the Identification of Technical Debt Through Code Comment Analysis. Lecture Notes in Business Information Processing. pp. 284-309, 2017. 
\item \label{SP29} 
M. Ciolkowsk, L. Guzmán, A. Trendowicz and F. Salfner. Lessons Learned from the ProDebt Research Project on Planning Technical Debt Strategically. International Conference on Product-Focused Software Process Improvement. pp. 523-534, 2017.
\item \label{SP30} 
H. Ghanbari, T. Besker, A. Martini and J. Bosch. Looking for Peace of Mind? Manage Your (Technical) Debt: An Exploratory Field Study. International Symposium on Empirical Software Engineering and Measurement (ESEM), pp. 384-393, 2017.
\item \label{SP31} 
A. Martini and J. Bosch. Revealing Social Debt with the CAFFEA Framework: An Antidote to Architectural Debt.  International Conference on Software Architecture Workshops (ICSAW),pp. 179-181, 2017
\item \label{SP32} 
A. Martini, S. Vajda, J. Vasa, A. Jones, M. Abdelrazek, J. Grundy and J. Bosch. Technical debt interest assessment: from issues to project.  XP2017 Scientific Workshops. pp. 1-6, 2017. 
\item \label{SP33} 
A. Martini and J. Bosch. The magnificent seven: towards a systematic estimation of technical debt interest. IXP2017 Scientific Workshops (XP '17), Article 7, 5 pages, 2017.
\item \label{SP34} 
T. Besker, A. Martini and J. Bosch. The Pricey Bill of Technical Debt: When and by Whom will it be Paid?.  International Conference on Software Maintenance and Evolution (ICSME), pp. 13-23, 2017
\item \label{SP35} 
A.  Martini, E. Sikander, and N. Madlani. A semi-automated framework for the identification and estimation of Architectural Technical Debt. Information and  Software Technology, Vol. 93, C, pp.264-279, 2018.
\item \label{SP36} 
J. M. Conejero, R. Rodríguez-Echeverría, J. Hernández, P. J. Clemente, C. Ortiz-Caraballo, E. Jurado, F. Sánchez-Figueroa, Early evaluation of technical debt impact on maintainability. Journal of Systems and Software. Vol. 142, pp. 92-114, 2018.
\item \label{SP37} 
A. Martini, T. Besker, J. Bosch. Technical Debt tracking: Current state of practice: A survey and multiple case study in 15 large organizations. Science of Computer Programming. Vol. 163, pp. 42-61, 2018. 
\item \label{SP38} 
A. Martini, F. Arcelli Fontana, A. Biaggi, R. Roveda. Identifying and Prioritizing Architectural Debt Through Architectural Smells: A Case Study in a Large Software Company. 12th European Conference on Software Architecture (ECSA), pp. 24-28, 2018
\item \label{SP39} 
A. Tornhill. Prioritize technical debt in large-scale systems using codescene. International Conference on Technical Debt (TechDebt ’18), pp. 59–60, 2018
\item \label{SP40} 
M. Albarak and R. Bahsoon. Prioritizing technical debt in database normalization using portfolio theory and data quality metrics. International Conference on Technical Debt (TechDebt ’18), pp. 31–40, 2018
\item \label{SP41} 
H.M. Firdaus and H. Lichter. Towards a Technical Debt Management Framework based on Cost-Benefit Analysis. ICSEA 2015, 2015
\item \label{SP42} 
R. Pl{\"o}sch, J. Br{\"a}uer, M. Saft, and C. K{\"o}rner. Design debt prioritization: a design best practice-based approach. International Conference on Technical Debt (TechDebt ’18), pp. 95–104, 2018
\item \label{SP43} 
R. Rebouças de Almeida, U. Kulesza, C. Treude, D. Cavalcanti Feitosa and A. Higino Guedes Lima. Aligning Technical Debt Prioritization with Business Objectives: A Multiple-Case Study. International Conference on Software Maintenance and Evolution (ICSME 2018), pp. 655-664, 2018
\item \label{SP44} 
R. Alfayez and B. Boehm. Technical Debt Prioritization: A Search-Based Approach. 19th International Conference on Software Quality, Reliability and Security (QRS),  pp. 434-445, 2019
\end{enumerate}

\end{document}